\documentclass[format=acmsmall, review=false, screen=true]{acmart}

\usepackage{booktabs} % For formal tables

\usepackage[ruled]{algorithm2e} % For algorithms

\usepackage{algorithmic}
\usepackage{algorithm2e}
\SetAlFnt{\small}
\SetAlCapFnt{\small}
\SetAlCapNameFnt{\small}
\SetAlCapHSkip{0pt}
\IncMargin{-\parindent}

\usepackage{helvet}  %Required
\usepackage{courier}  %Required
\usepackage{url}  %Required
\usepackage{graphicx}  %Required
\usepackage{booktabs}
\usepackage{amsmath}

\usepackage{multirow}
\usepackage{multirow}
\usepackage{makecell}
\usepackage{caption}
\usepackage{subfig}

\newcommand{\tabincell}[2]{\begin{tabular}{@{}#1@{}}#2\end{tabular}} 
\begin{document}
% Title portion. Note the short title for running heads
\title{Cascading: Association Augmented Sequential Recommendation}

\author{Xu Chen}
\affiliation{%
  \institution{Cooperative Medianet Innovation Center, Shanghai Jiao Tong University}
  \streetaddress{800 Dongchuan Rd}
  \city{Shanghai}
  \country{China}}
\email{xuchen2016@sjtu.edu.cn}

\author{Kenan Cui}
\affiliation{%
  \institution{Cooperative Medianet Innovation Center, Shanghai Jiao Tong University}
  \city{Shanghai}
  \country{China}}
\email{conancui@sjtu.edu.cn}

\author{Ya Zhang}
\affiliation{%
  \institution{Cooperative Medianet Innovation Center, Shanghai Jiao Tong University}
  \city{Shanghai}
  \country{China}}
\email{ya_zhang@sjtu.edu.cn}

\author{Yanfeng Wang}
\affiliation{%
  \institution{Cooperative Medianet Innovation Center, Shanghai Jiao Tong University}
  \city{Shanghai}
  \country{China}}
\email{wangyanfeng@sjtu.edu.cn}

\begin{abstract}
Recently, recommendation according to sequential user behaviors has shown promising results in many application scenarios. Generally speaking, real-world sequential user behaviors usually reflect a hybrid of sequential influences and association relationships. However, most existing sequential recommendation methods mainly concentrate on sequential relationships while ignoring association relationships. %Recent sequential behavior modeling methods (e.g. RNNs) have
%for sequential recommendation mainly concentrate on successfully captured the sequential relationships but ignored item association relationships. 
In this paper, we propose a unified method that incorporates item association and sequential relationships for sequential recommendation.
Specifically, we encode the item association as relations in item co-occurrence graph and mine it through graph embedding by GCNs.  In the meanwhile, we model the sequential relationships through a widely used RNNs based sequential recommendation method named GRU4Rec. The two parts are connected into an end-to-end network with cascading style, which guarantees that representations for item associations and sequential relationships are learned simultaneously and make the learning process maintain low complexity. We perform extensive experiments on three widely used real-world datasets: TaoBao, Amazon Instant Video and Amazon Cell Phone and Accessories. Comprehensive results have shown that the proposed method outperforms several state-of-the-art methods. Furthermore, a qualitative analysis is also provided to encourage a better understanding about association relationships in sequential recommendation and illustrate the performance gain is exactly from item association.
\end{abstract}





\maketitle

\renewcommand{\shortauthors}{X. Chen et al.}

\section{Introduction}
Recommendation, as an information filtering method, has been extended to a wide range of real-world applications such as product recommendation~\cite{Qu:2018:PNN:3289475.3233770}, video recommendation~\cite{Lu:2018:DBT:3289475.3233773} and app recommendation~\cite{Yao:2017:VRP:3112649.3015458}. In recent years, an important trend in recommendation is to consider the order of user behaviors~\cite{Hidasi2015Session,Smola2017Neural,Wu2017Recurrent,Yu2016A}, which has shown promising results by capturing behavior relationships. According to the way to incorporate sequential information, existing methods can be summarized into two categories. Traditional methods utilize sequential information as contextual features during training~\cite{Gao2013Exploring,Koenigstein2011Yahoo,Yuan2013Time}, which cannot model the high-order sequential relationships. Recent sequential recommendation methods inherently incorporate temporal information with Recurrent Neural Networks (RNNs)~\cite{Hidasi2015Session,Wu2017Recurrent,Yu2016A}. And these methods are also explored by introducing attention mechanism~\cite{Donkers2017Sequential,Pei2017Interacting}, personalization~\cite{Donkers2017Sequential} and auxiliary information (e.g. heterogeneous attributes, knowledge base)~\cite{liu2018sequential, huang2018improving, chen2018sequential} in order to model sequential user behavior better.

\begin{figure}
    \centering
    \includegraphics[width=3.5in]{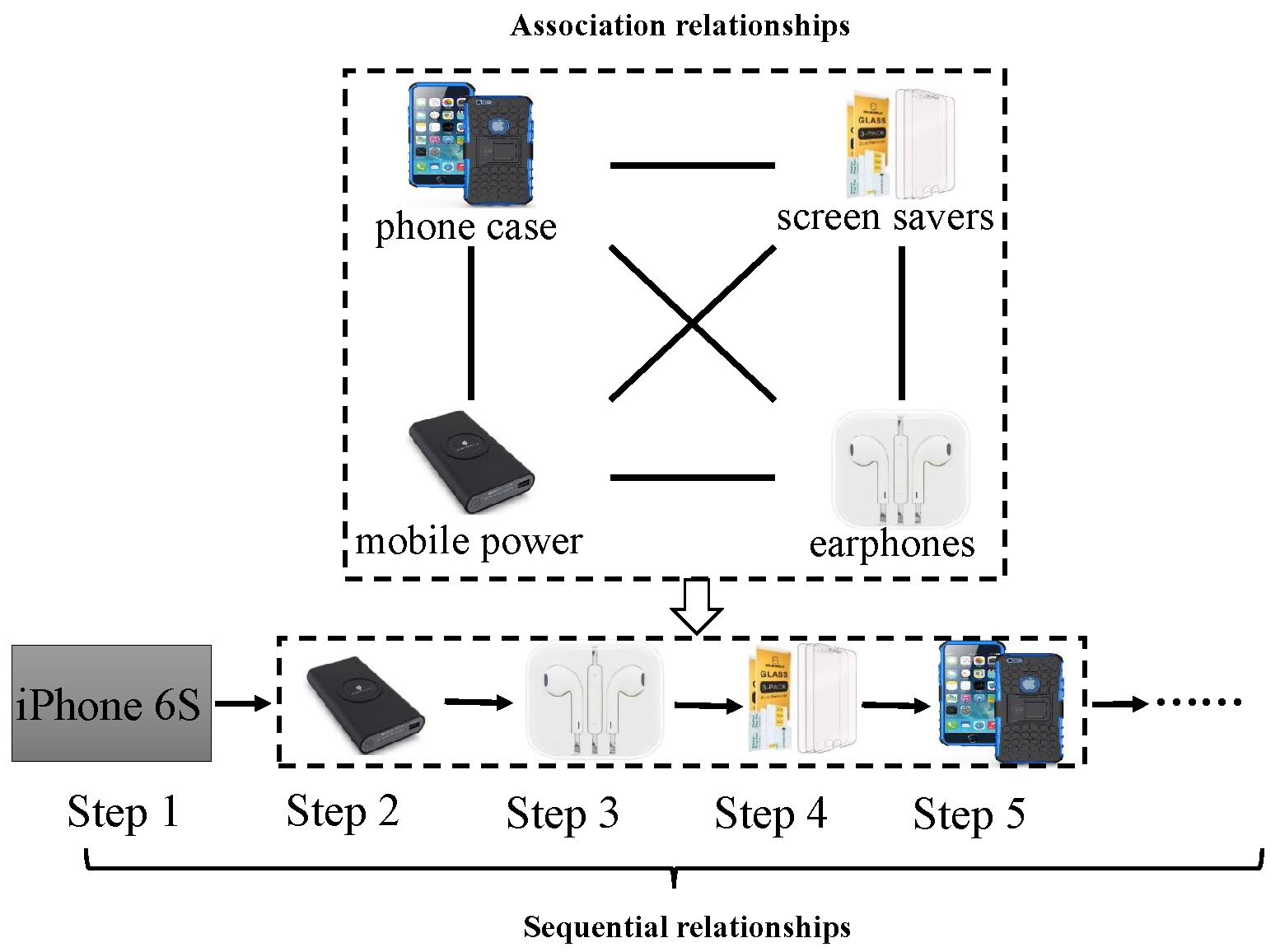}
    \caption{An example to illustrate the sequential relationships and association relationships in consuming sequence. The directed edge among items indicates the sequential relationships and undirected edge means the association relationships.}
    \label{figure:motivation}
\end{figure}

However, the real-world user behaviors usually reflect a hybrid of sequential relationships and association relationships. The orders observed in behavior sequences do not necessarily reflect relationships among behaviors. For example, Figure~\ref{figure:motivation} illustrates an user's purchase sequence of phone and accessories. Clearly, the transition from iPhone 6s (step 1) to phone accessories (step 2,3,4,5) contains sequential relationships due to purchase causality. On the other hand, the purchase order of cellphone accessories in fact could be arbitrary and actually reflects their association. 

\begin{figure}
    \centering
    \includegraphics[width=3.5in]{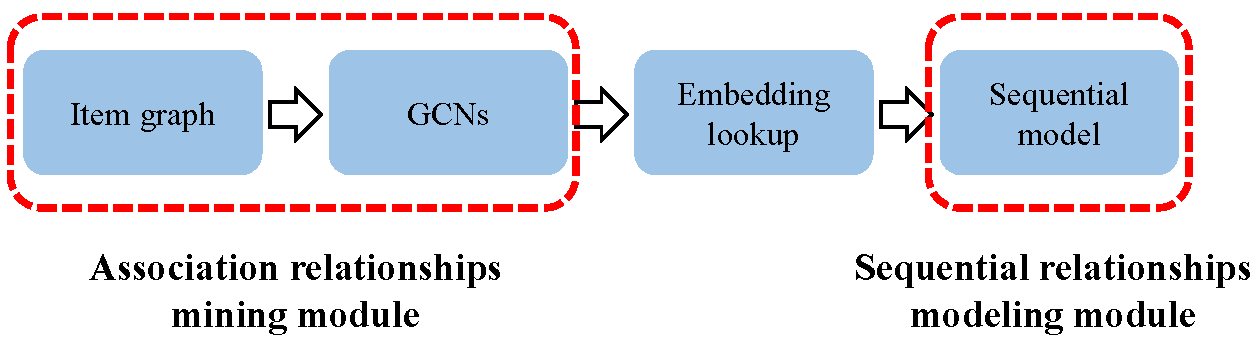}
    \caption{The general architecture of CAASR. The item graph and GCNs aim to mine the item association relationships and the sequential model aims to capture the sequential relationships in sequential user behavior.}
    \label{figure:model_general}
\end{figure}

Although RNNs based sequential modeling methods are capable of capturing the sequential relationships among user behaviors, they ignore the association relationships among items. 
%This happens mainly due to the fact that RNNs aims to model the sequential relationships explicitly but does not put explicit learning paradigm for association relationships. 
We here thus propose an unified model named CAASR (Cascading: Association Augmented Sequential Recommendation) to simultaneously capture association relationships and sequential relationships for sequential recommendation. 
The general idea of CAASR is illustrated in Figure~\ref{figure:model_general}. The item association features are mined from item graph by GCNs, and the sequential relationships are modeled by a widely used RNNs based sequential Recommendation method named GRU4Rec~\cite{Hidasi2015Session}. The two are connected in cascades and learned with one single training target. 
Because the adopted GCNs does not support mini-batch training, we design a specific graph embedding lookup layer which can combine GCNs and RNNs based sequential modeling method adaptively.
Table~\ref{table:methods_comparison} compares CAASR with other methods in various perspectives. The CAASR is the only method that mines item association and generalizes RNNs based sequential recommendation. 

% Stacking previous analysis together, we propose an end-to-end association augmented recommendation method for modeling sequential user behavior data in cascading style, Cascading: Association Augmented Sequential Recommendation (CAASR). 
% The general idea of CAASR is illustrated in Figure~\ref{figure:model_general}. In CAASR, we mine the item association features from item graph by GCNs and model the sequential relationships by a widely used RNNs based sequential recommendation method named GRU4Rec~\cite{Hidasi2015Session}. CAASR works in cascading style which guarantees knowledgeable item features for sequential recommendation module and in turn the sequential recommendation module further refines features learned from item graph. In other words, CAASR is learned with one training target, making the model share both item association relationships and sequential information flow. Note that the adopted GCNs does not support mini-batch training which is widely used in many deep learning methods. We thus design a specific graph embedding lookup layer which can combine GCNs and RNNs based sequential modeling method adaptively. Table~\ref{table:methods_comparison} compares CAASR with other methods in various perspectives. The cascading CAASR is the only method that mines item association and generalizes RNNs based sequential recommendation. 
\begin{table}[]
\caption{Comparison of CAASR with other methods in different perspectives. P-Cofactor and P-GraphAE both are methods of capturing item association relationships in parallel style. Specifically, P-Cofactor adopts co-factorization technique in ~\cite{liang2016factorization}, while P-GraphAE utilizes widely used Autoencoder regularization technique in~\cite{cao2017embedding}. CAASR is our proposed method in cascading style.}
\label{table:methods_comparison}
\begin{tabular}{ccccc}
\hline
Method     & \begin{tabular}[c]{@{}c@{}}sequential \\relationships\end{tabular}& \begin{tabular}[c]{@{}c@{}}item \\association \end{tabular}& \begin{tabular}[c]{@{}c@{}}association\\ combination style\end{tabular} & \begin{tabular}[c]{@{}c@{}}generalize\\ RNNs based\\ sequential recommendation\end{tabular} \\ \hline
BPR        & $\times$              & $\times$               & -                                                                       & $\times$                                                                              \\
BPR+KNN    & $\times$              & $\times$               & -                                                                       & $\times$                                                                              \\
GRU4Rec    & \checkmark             & $\times$               & -                                                                       & $\times$                                                                              \\
P-Cofactor & \checkmark             & \checkmark              & parallel                                                                & $\times$                                                                              \\
P-GraphAE  & \checkmark             & \checkmark              & parallel                                                                & $\times$                                                                              \\
CAASR      & \checkmark             & \checkmark              & cascading                                                               & \checkmark                                                                             \\ \hline
\end{tabular}
\end{table}

Our contributions are summarized as three following points:
\begin{itemize}
\item In this paper, we firstly concentrate on modeling both the item association relationships and sequential relationships for sequential recommendation, and develop a novel method named CAASR. To the best we know, CAASR is the first model that concentrates on item association relationships in RNNs based sequential recommendation.
    
\item To seamlessly combine item association relationships and sequential relationships, we explore a better cascading style to incorporate them rather than the widely used parallel regularization technique. 
    
\item We conduct extensive experiments on three real-world datasets. Comprehensive results demonstrate the superiority over state-of-the-art models both quantitatively and qualitatively.
\end{itemize}

The rest of this paper is organized as follows. Section 2 introduces recent related works, including sequential recommendation, item association relationships in recommendation and graph embedding methods. Problem definition and a concise introduction of GCNs are given in section 3. Section 4 provides the details about proposed methods. Experiments and results are illustrated in section 5, and the results of our algorithm are verified both quantitatively and qualitatively. Finally, conclusion and future work are given in section 6.  

\section{Related Work}
\subsection{Towards Recommendation Methods for Sequential User Behavior}
Most sequential recommendation methods are designed to capture and utilize the sequential relationships from sequence data. 

\textbf{Traditional Methods}. Previously, sequential information is usually used as a contextual feature during the training phase, in which timestamp is considered as an additional resource to enrich the model features. However, these methods~\cite{Karatzoglou2010Multiverse,Gao2013Exploring} based on handcrafted features are consuming and not applicable. Also TimeSVD++~\cite{koren2009collaborative} tends to exploit temporal signals through SVD technique. Tensor factorization is another way of applying temporal information such as~\cite{Li:2017:TPP:3112649.3057283}. They typically cannot model high-order sequential relationships from chronological data~\cite{Donkers2017Sequential}. There are also several sequential models often decomposing the problem into two parts: user preference modeling and sequential modeling. For example, Factorized Personalized Markov Chain (FPMC)~\cite{rendle2010factorizing} is a classic sequential recommendation that fuses MF and factorized Markov Chain to model user preference and sequential relationships respectively. Recently, inspired by transnational metric embeddings~\cite{bordes2013translating}, TransRec~\cite{he2017translation} unifies sequential recommendation and a specific metric embedding approach together, modeling each user as a translating vector from his/her last visited item to the next one.

\textbf{Deep Learning Based Methods}. Recently, deep learning methods present excellent representation learning ability, like Convolutional Neural Network (CNNs)~\cite{Ketkar2017Convolutional} and Recurrent Neural Networks (RNNs)~\cite{Rumelhart1986Learning}. RNNs is one promising approach in capturing the sequential relationships of user preference via user-item interaction data~\cite{Smola2017Neural,Wu2017Recurrent,Yu2016A}. Wu et.al~\cite{Wu2017Recurrent} and Hidasi et.al~\cite{Hidasi2015Session} utilize Long Short Term Memory (LSTM)~\cite{Graves2012Long} and Gated Recurrent Unit (GRU)~\cite{Cho2014On} respectively to model sequential relationships of user representation from chronological data. Then, in~\cite{Quadrana2016Parallel}, both user representation and item one-hot vectors are merged into one model based on~\cite{Hidasi2015Session}. Tim et.al~\cite{Donkers2017Sequential} point that approaches in~\cite{Hidasi2015Session,Quadrana2016Parallel} do not explicitly model individual users, thus they introduce a variant of GRU model that utilizes pair-wise loss to generate personalized recommendation. Due to the well performance of attention mechanism in Neural Machine Translation (NMT)~\cite{Bahdanau2014Neural} and video captioning tasks~\cite{Chen2016SCA,Xu2015Show}, attention mechanism is introduced into recommendation as well in~\cite{Donkers2017Sequential,Pei2017Interacting,li2017neural,liu2018stamp}. In~\cite{Donkers2017Sequential}, they design a kind of attention gated cell to regulate the original gating process in GRU. Interacting Attention-gated Recurrent Networks (IARN), proposed in~\cite{Pei2017Interacting}, extends recurrent networks for both modeling user and item dynamics with a novel gating mechanism. A novel attention scheme is also designed to allow the user-side and item-side recurrent networks to interact with each other. STAMP~\cite{liu2018stamp}, a novel short-term attention/memory priority mechanism is proposed to emphasize general interests of users. As for auxiliary information, item heterogeneous attributes and knowledge base has been applied in sequential recommendation in~\cite{liu2018sequential} and~\cite{huang2018improving, chen2018sequential} respectively.

\subsection{Item Association in Recommendation}
Although item association is rarely modeled in RNNs based sequential recommendation, it has been widely used in non-sequential recommendation methods~\cite{Mei:2011:CVR:1961209.1961213,liang2016factorization,cao2017embedding}. VideoReach in~\cite{Mei:2011:CVR:1961209.1961213} presents a better video recommendation method based on multi-modal content relevance. While in many recommendation scenarios, contents of items are difficult to obtain. In item-based collaborative filtering, an item-item similarity matrix encoding item relevance is defined from user behavior data to directly apply the similarity matrix to predict missing preferences. This method is not practical and highly sensitive to the choice of similarity metric and data normalization~\cite{herlocker2002empirical}. While inspired by the exploration of co-factorizing multiple relational matrices in Collective Matrix Factorization (CMF)~\cite{singh2008relational}, Cofactor~\cite{liang2016factorization} constructs \textit{SPPMI} matrix from item co-occurrence matrix and applies co-factorization as the regularization term of item representation in Matrix Factorization (MF) method~\cite{koren2009matrix}. This co-factorization technique is also applied into ranking recommendation in~\cite{cao2017embedding}. 
% Regarding the deep representation of user-item interaction~\cite{he2017neural, cheng2016wide, zhou2018deep}, Bai et.al~\cite{bai2017neural} utilize KNN technique to model item relation into these deep models.

\subsection{Graph Embedding Methods}
Representation learning about nodes in graphs has arisen as one hot research area. These methods mainly focus on two kinds of directions. 

\textbf{Random walk Based}. Recently, skip-gram model~\cite{mikolov2013efficient} has shown its success in natural language processing. This method has opened new ways for feature learning of discrete objects such as words. Perozzi et.al~\cite{perozzi2014deepwalk} discover that random walks of nodes in graph lead to similar power-law distribution like words in natural language processing. Thus they regard the random walks on graph as sentences and leverage skip-gram model for learning latent node representation. In~\cite{tang2015line}, Tang et.al define a node's context by its neighborhoods. Node2Vec~\cite{grover2016node2vec} defines biased random walk to control the Bread First Search (BFS) and Deep First Search (DFS). These random walk motivated works are based on skip-gram and Levy et.al~\cite{levy2014neural} point that skip-gram with negative sampling (SGNS)~\cite{mikolov2013efficient} is actually conducting implicit matrix factorization.

\textbf{Graph Convolutional Networks}. Graph Convolutional Networks is the other representative methodology for this research area from spectral graph analysis. 

In spectral graph theory~\cite{Chung1997Spectral}, complex geometric structures in graphs can be studied with spectral filters. Graph Convolutional Network learns graph structure from a spectral graph view. It is a kind of network that learns local stationary features of nodes in spectral graph domain. 

In the convolution theorem~\cite{mallat1999wavelet}, convolution is defined as linear operators that diagonalized in Fourier basis. And this basis is the eigenvectors of graph Laplacian operator. While this kind of filter defined in spectral domain cannot naturally be localized and the computation is costly because of multiplication with graph Fourier. These limitations are overcome by a special choice of filter parameterization in~\cite{Bruna2013Spectral,Defferrard2016Convolutional}. In~\cite{Defferrard2016Convolutional}, the authors apply Chebyshev polynomials to reach a recursive formulation of graph convolution operation. Compared with work in~\cite{Bruna2013Spectral}, method in~\cite{Defferrard2016Convolutional} provides a strict control over local support of filters and is more computationally efficient by avoiding using the eigenvalue decomposition of graph Laplacian matrix. Meanwhile, better performance is achieved as well. GCNs is similar to convolution operation in CNNs, while this is in spectral domain for graph data. Kipf et.al~\cite{Kipf2016Semi} simplifies GCNs in~\cite{Defferrard2016Convolutional} through stacking convolutional layers instead of summation of different Chebshev orders. Velickovic et.al~\cite{velickovic2017graph} regard graph convolution as the aggregation problem of node representation and they introduce multi-head attention mechanism to learn attentive weights for each nodes during aggregation process. GraphSAGE in~\cite{hamilton2017inductive} generalizes graph convolution network to unseen nodes and proposes an inductive learning framework that leverages node feature information to efficiently generate node embeddings. Note that we mainly focus on~\cite{Defferrard2016Convolutional} here. A concise introduction of GCNs exploited in our paper would be provided later.

\section{Preliminary}
\subsection{Problem Definition}
We firstly introduce the notations utilized in our paper and give a clear problem definition in recommendation for sequential user behavior data. In our paper, we denote $\mathcal{U}$ as the set of users and $\mathcal{I}$ as the set of items. Our task concentrates on implicit feedback in user behavior data, where feedback between user $u\in \mathcal{U}$ and item $i\in \mathcal{I}$ at time step $t$ is denoted as 1 if they have interaction or 0 if not. By ascendingly sorting the interaction data of each user according to time, we can form the sequence data for the user as $\{i_{u}^{1},...,i_{u}^{t},...,i_{u}^{n_{u}}\}$, where $i_{u}^{t}$ represents the item that user $u$ has interacted with at time step $t$ and $n_{u}$ is the length of interaction data for user $u$. Given an interaction sequence $\{i_{u}^{1},...,i_{u}^{t}\}$, our aim lies in predicting the top $M$ items that the user would most probably interact with at time step $t+1$. And the notations utilized in this paper are summarized in Table~\ref{table:notations}.
\begin{table}[]
\caption{Notations in this paper.}
\label{table:notations}
\begin{tabular}{ll}
\hline
Notation & Description \\ \hline
$\mathcal{U}$        & the user set            \\
$\mathcal{I}$        & the item set            \\
$N$             & number of users (sequences)   \\
$M$            & number of items  \\
$i_{u}^{t}$    & the item that user $u$ interacted at step $t$ \\
$A$  & item graph adjacent matrix            \\
$D$    & degree matrix of $A$ \\
$I_{N}$        & the identity matrix of size $N$ \\
$L$        & symmetric normalized Laplacian matrix of item graph         \\
$\widetilde{L}$ & rescaled symmetric normalized Laplacian matrix of item graph \\
$K$        & Chebshev order of graph convolution in this paper\\
$d$        & dimension of embedding \\ 
$T_{k}(x)$       & Chenshev polynomial with order $k$           \\
$L_{CAASR}$                  & loss function of CAASR method           \\ 
$L_{s}$          & loss function of pure RNNs based sequential recommendation method               \\
$S$              & \textit{SPPMI} matrix              \\
$L_{P-Cofactor}$     & loss function of P-Cofactor method          \\
$L_{P-GraphAE}$    & loss function of P-GraphAE method   \\
\hline
\end{tabular}
\end{table}
\subsection{Graph Convolutional Networks}
Graph Convolutional Networks (GCNs) is one essential ingredient for our CAASR method, thus we give a concise introduction about it. GCNs is an operation that aims to learn superior node representation considering dependence among nodes. It is neither limited to the task nor the model~\cite{Defferrard2016Convolutional,Kipf2016Semi,velickovic2017graph,hamilton2017inductive}.

Spectral convolution on graphs is defined as the multiplication of a signal $x\in R^{N}$ with a parameterized filter $g_{\theta}$ in the Fourier domain, i.e.:
\begin{equation}
    g_{\theta}\star x=Ug_{\theta}(\Lambda)U^{T}x
\end{equation}
where $\star$ represents the graph convolution operation. $U$ and $\Lambda$ denote the matrix of eigenvectors and eigenvalues of the graph Laplacian $L=I_{N}-D^{-\frac{1}{2}}AD^{-\frac{1}{2}}=U\Lambda U^{T}$, respectively. And $U^{T}x$ indicates the graph Fourier transform of $x$. And a polynomial filter is taken in~\cite{Defferrard2016Convolutional} as $ g_{\theta}(\Lambda)=\sum_{k=0}^{K}\theta_{k}\Lambda^{k}$. While this convolution filter involves the eigen-decomposition of $L$ which might be computationally expensive for large graphs.  To circumvent this problem, $g_{\theta}(\Lambda)$ could be well-approximated by a truncated expansion in terms of Chebshev polynomials $T_{k}(x)$ up to $K^{th}$ order~\cite{hammond2011wavelets}:
\begin{equation}
    g_{\theta}(\Lambda)\approx \sum_{k=0}^{K}\theta_{k}T_{k}(\widetilde{\Lambda}) \Rightarrow g_{\theta}\star x \approx \sum_{k=0}^{K}\theta_{k}T_{k}(\widetilde{L})x
\label{eq:graph_convolution_chebshev}
\end{equation}
where $\widetilde{\Lambda}=\frac{2}{\lambda_{max}}\Lambda-I_{N}$ and $\widetilde{L}=\frac{2}{\lambda_{max}}L-I_{N}$. $\lambda_{max}$ denotes the largest eigenvalue of $L$. The Chebshev polynomails are recursively defined as $T_{k}(x)=2xT_{k-1}(x)-T_{k-2}(x)$ with $T_{0}=1$ and $T_{1}=x$. 

From above we could clearly see that GCNs learn each node representation from the spectral graph domain. The size of one node's neighbors is called the \textit{receptive field}. It is enlarged through increasing Chebshev order $K$ just like the $L$-hop in a graph, which could encourage more neighborhood information when learning node representation.
\begin{figure*}[h!]
\centering
\begin{minipage}[t]{0.4\textwidth}
\centering
\includegraphics[width=\textwidth]{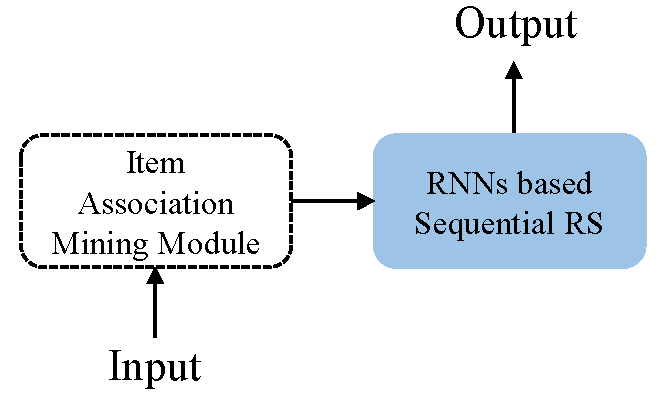}
\vspace{-20pt}
\caption*{(a) cascading style}
\end{minipage}
\begin{minipage}[t]{0.4\textwidth}
\centering
\includegraphics[width=\textwidth]{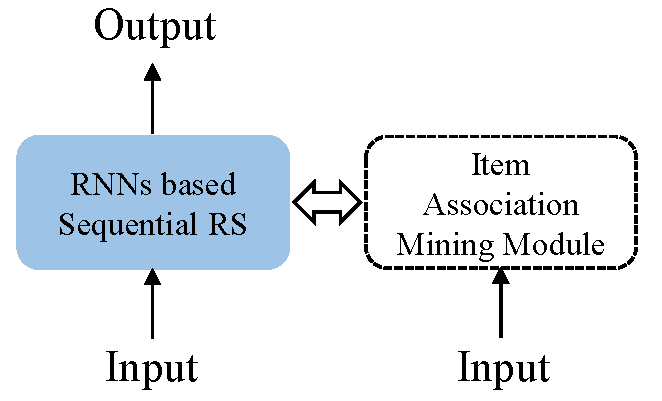}
\vspace{-20pt}
\caption*{(b) parallel style}
\end{minipage} \\
\caption{In sequential user behavior, combine item association relationships and sequential relationships in two kinds of styles. Item association mining module aims to model the item association relationships and the RNNs based sequential recommendation aims to capture the sequential relationships.}
\label{figure:combine_style}
\end{figure*}

\section{The Proposed Method}
\begin{figure*}
    \centering
    \includegraphics[width=5.5in]{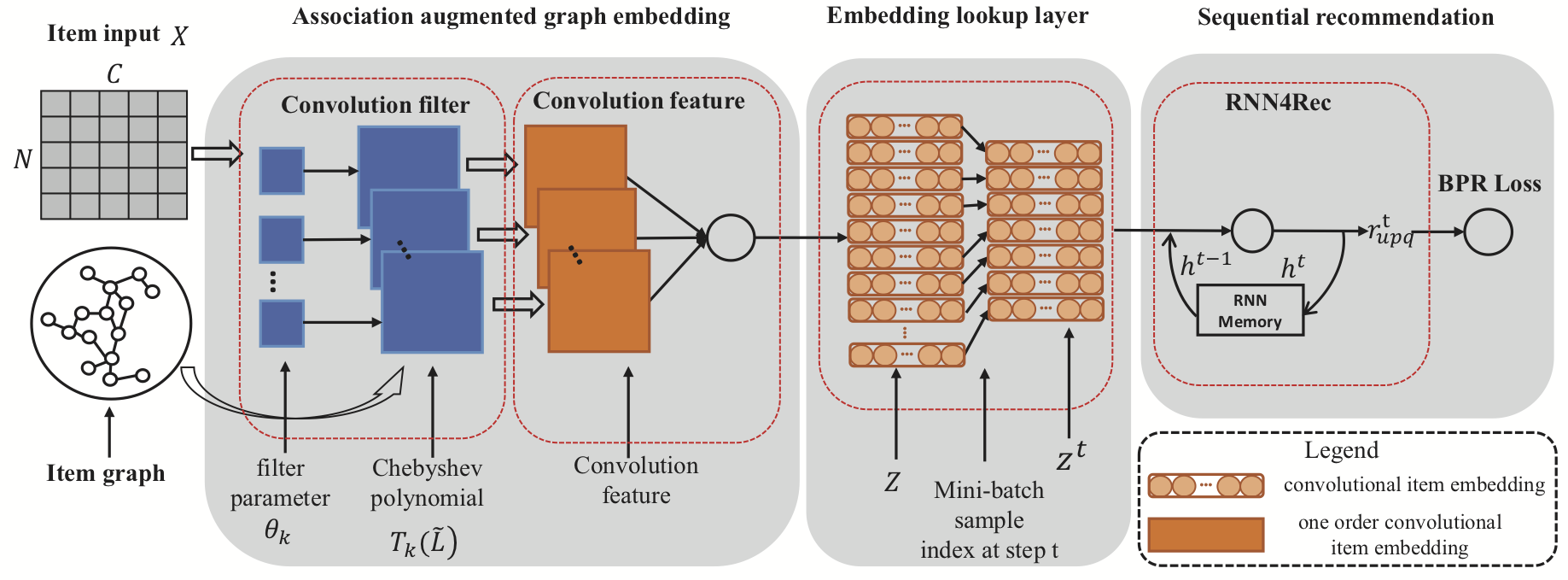}
    \caption{The architecture of our proposed CAASR model. The pngcascading procedure is well designed by three components: Association augmented graph embedding, Embedding lookup layer and Sequential recommendation.}
    \label{figure:model_architecture}
\end{figure*}
In this section, we demonstrate: 1) how to sufficiently mine item association relationships merely from user behavior data; 2) how to combine this item association relationships with sequential relationships in different styles. Both cascading and parallel style are explored and shown in Figure~\ref{figure:combine_style}.

We firstly introduce the general architecture of our proposed cascading CAASR model illustrated in Figure \ref{figure:model_architecture}. Apart from this, two parallel styles of combining this item association pattern are also explored and analyzed later. Details are illustrated as follows.
\subsection{Association Augmented Graph Embedding}
\label{sec:ASSR}
As we summarized before, the real-world user consuming sequence is usually a hybrid of association relationships and sequential relationships. And recent RNNs based sequential recommendation methods such as GRU4Rec are capable of capturing the sequential relationships but fail to emphasize the item association characteristic. We expect to model this item association pattern while performing sequential recommendation task.

As for mining the item association relationships, there are several existing non-sequential recommendation methods and the typical one is~\cite{liang2016factorization}. In~\cite{liang2016factorization}, an item co-occurrence matrix is constructed from user behavior data and co-factorization technique is utilized. Whereas we find this technique does not work well in RNNs based sequential user behavior modeling. This happens mainly because it is incompatible to apply $L_{2}$ regularization between deeply represented representation by RNNs and shallowly learned representation by traditional matrix factorization. Instead, as one kind of expression style of graph data, this co-occurrence matrix intrinsically reflects association relationship among items. In order to extract item association information, we use item co-occurrence data to construct the item graph which could be easily processed by graph embedding methods.

Considering graph embedding methods, they can be categorized into two main groups, the random walk based and GCNs. The random walk based graph embedding methods do not support end-to-end training as well as rely a lot on the quality of random walk sequences. In contrast, GCNs, developed from spectral graph signal processing~\cite{shuman2013emerging}, has robust and superior performance over random walk based methods, which has been proved in~\cite{Defferrard2016Convolutional, Kipf2016Semi}. It is neither limited to the model nor the task, allowing an end-to-end and flexible combination for sequential recommendation. Therefore, we leverage graph convolution on the item co-occurrence graph to learn association augmented item embeddings as prior knowledge for describing items. 

\textbf{Graph Construction}
\begin{figure*}
    \centering
    \includegraphics[width=4.5in]{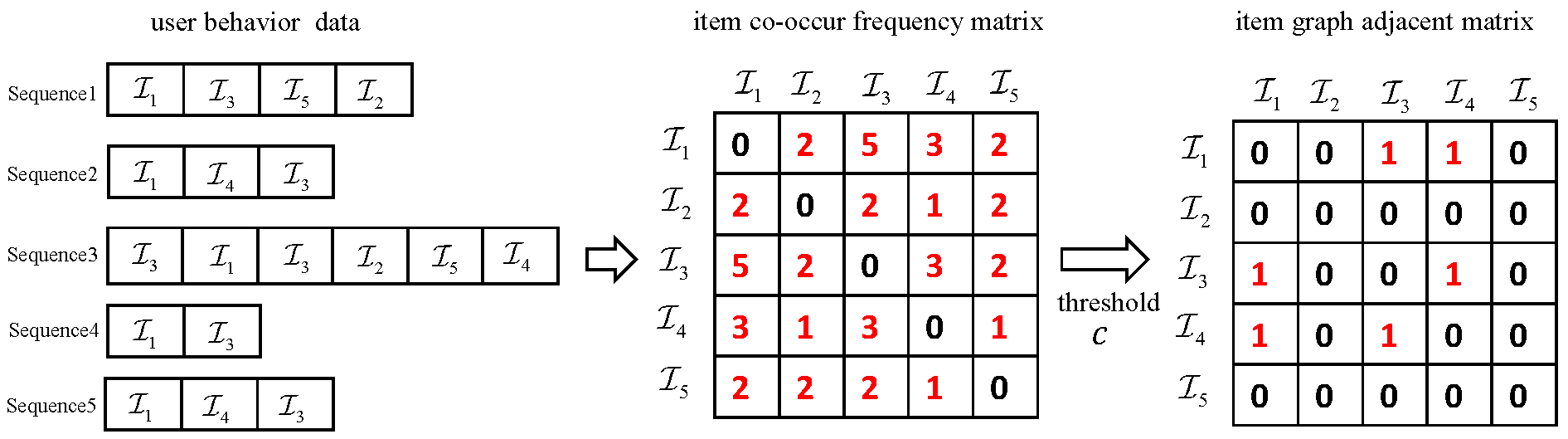}
    \caption{An example to illustrate the item graph construction process. The threshold $c$ is set to 2 in this example.}
    \label{figure:item_graph_construction}
\end{figure*}
The item co-occurrence graph can be obtained from user-item interactions. We count the number of user consuming sequence that two items $\mathcal{I}_{i},\mathcal{I}_{j}$ co-occur. By this way, a matrix containing the frequency that every two items co-occur is formed. We set the frequency number larger than a threshold $c$ to 1 and the rest as 0 to form a sparse adjacent matrix $A$ of the item graph. This adjacent matrix $A\in R^{N\times N}$($N$ is the number of items) is symmetric here, indicting the symmetric relationship between every two items. We give an example to illustrate this process in Figure~\ref{figure:item_graph_construction}. 

To prepare for the subsequent convolution operation in spectral domain, several operations need to be performed according to graph convolutional theory in~\cite{Defferrard2016Convolutional}. Firstly, we obtain the normalized graph Laplacian $L$ through $L=I_{N}-D^{-\frac{1}{2}}AD^{-\frac{1}{2}}$, where $D\in R^{N\times N}$ is the diagonal degree matrix with $D_{ii}=\sum_{j}A_{ij}$ and $I_N$ is the identity matrix of size $N$. Then, the re-scaled version of $L$ is formulated as $\tilde{L}=\frac{2}{\lambda_{max}}L-I_N$ and $\lambda_{max}$ is the maximum eigenvalue of $L$.  

\textbf{Graph Convolution}. In order to extract item association information from item graph $A$, this item graph needs to be filtered in spectral domain. Suppose we have the item representation input $X\in R^{N\times C}$, $N$ is the number of items and $C$ is the input channel for each item. Here we only use the one-hot vector representation for all items as $X$ in all our model training process, this means $X=I_N$ and $I_N$ is an identity matrix with size $N$. The convolution on item graph can be generalized as below:
\begin{equation}
\label{equation:X_convolution}
Z=g_{\theta} \star X \approx \sum_{k=0}^{K}T_{k}(\tilde{L})X\Theta_k
\end{equation}
where $\star$ indicates the graph convolution operation defined in spectral domain. $K$ is the number of order of Chebyshev polynomials in Laplacian. $\Theta_k\in R^{C\times d}$ is the filter parameter that needs to be learned for order $k$. $Z\in R^{N\times d}$ denotes the spectral graph embeddings of all items, and it contains all the associated structure information of items. Specifically, the Chebyshev polynomial $T_{k}(\tilde{L})$ is a sparse matrix, which can be recursively generated by using the definition in Eq.~\ref{equation:chebyshev polynomials}. And $T_{k}(\tilde{L})X\Theta_{k}$ can be efficiently implemented as the product of a sparse matrix with a dense matrix by using Tensorflow\footnote{https://www.tensorflow.org/} or Pytorch\footnote{https://pytorch.org/}.
\begin{equation}
\begin{split}
&T_{k}(\tilde{L})=2\tilde{L}T_{k-1}(\tilde{L})-T_{k-2}(\tilde{L}) \\
& T_{0}(\tilde{L})=1, T_{1}(\tilde{L})=\tilde{L}
\end{split}
\label{equation:chebyshev polynomials}
\end{equation}

This association augmented graph embedding operation is illustrated is Figure~\ref{figure:model_architecture} as the first component in shade. The input $X$ is filtered by $K$ filters and outputs $K$ feature matrices each with shape $N\times d$ that capture different orders of associated features. Summarization of $K$ order features leads to the final convolutional features of all items $Z\in R^{N\times d}$.
        
\subsection{Embedding Lookup Layer}
\label{sec:emebdding_lookup_layer}
Through the spectral graph filtering, we get associated embeddings $Z$ of items. Note that in most neural networks, mini-batch training approach is utilized due to both low memory usage and better model stability. As for graph convolution in~\cite{Defferrard2016Convolutional,Kipf2016Semi}, mini-batch strategy cannot guarantee the original graph structure during training. Therefore we cannot train the spectral filter parameters $\Theta$ in mini-batch manner. Instead, we input the whole rescaled normalized graph Laplacian matrix $\tilde{L}$ in sparse manner. To seamlessly combine association augmented graph embedding and downstream sequential relationships modeling module with mini-batch training style, we design a graph embedding lookup layer to bridge the gap. In this manner, not only the graph structure is reserved but also the latter RNNs based sequential relationships modeling module can still be trained with mini-batch approach. 

This embedding lookup layer is shown in Figure~\ref{figure:model_architecture} as the second component with shade. In each forward and backward training process of final loss, considering we have a mini-batch item index that is $[1,3,4,5,7,8,N]$ at step $t$, then we lookup the corresponding association graph embeddings $Z$ that belong to items $[1,3,5,7,8,N]$. And the chosen mini-batch association item embedding at step $t$ is denoted as $z^{t}$. This means that in each epoch training process, one item graph embedding may be trained several times, which makes the learning process more stable and fast converged. And this derivable graph embedding lookup operation can be formulated as:
\begin{equation}
z^{t}=f_{lookup}(Z, \mathcal{I}_{index}^{t})
\end{equation}
where function $f_{lookup}(\cdot)$ denotes the lookup operation on association augmented graph embeddings $Z$. $z^{t}\in R^{s\times d}$ denotes the graph embedding of mini-batch with the corresponding item index $\mathcal{I}_{index}^{t}$, and $s$ is the mini-batch size.

\subsection{Sequential Recommendation}
\label{sec:sequential_recommendation}
Association augmented item embeddings from the former part is aimless. The latter sequential recommendation part captures the sequential relationships and refines associated embeddings for recommendation. This sequential recommendation part has been studied by many techniques such as attention mechanism~\cite{Donkers2017Sequential,Pei2017Interacting}, personalization~\cite{Donkers2017Sequential} and auxiliary information (e.g. heterogenous attributes, knowledge base)~\cite{liu2018sequential, huang2018improving, chen2018sequential}. While our model concentrates on modeling both item association relationships and sequential relationships integrally in sequential recommendation, thus we adopt one typical RNNs based sequential recommendation model named GRU4Rec~\cite{Hidasi2015Session} as the sequential relationships modeling module in our experiments\footnote{Note that this sequential recommendation module can be substituted with other RNNs based sequential models.}. GRU4rec is based on GRU which is a variant of RNN unit that aims to deal with the vanishing gradient problem. GRU4Rec also supports user cold-start problem due to the fact that it does not specify a user in training process, which is practical in real scenarios. The superior performance has pushed it as the state-of-art RNNs based sequential recommendation method. 

After we obtaining association augmented graph embeddings of items, we could sample the mini-batch spectral embeddings $z^t \in R^{s\times d}$ at step $t$.
Cascading manner allows the association information flow through GRU to be formulated as follows:

\noindent(1) the update gate:
\begin{equation}
c^t=\sigma(z^tW_c + h^{t-1}U_c)
\end{equation}
(2) the reset gate:
\begin{equation}
r^t=\sigma(z^tW_r + h^{t-1}U_r)
\end{equation}
(3) the candidate activation function $\hat{h}^t$ is:
\begin{equation}
\hat{h}^t=\tanh(z^tW_h + (r^t \odot h^{t-1})U_h)
\end{equation}
(4) the output state $h^t$ at step $t$ is given by:
\begin{equation}
h^t=(1-c^t)\odot h^{t-1}+c^t\odot \hat{h}^t
\end{equation}
where $W=\{W_c,W_r,W_z\} \in R^{3\times d\times d}$ and $U=\{U_c,U_r,U_z\} \in R^{3\times d\times d}$ are all parameters in GRU cell we need to learn. $\sigma$ denotes the $sigmoid$ activation function and $\odot$ is element-wise product.

\subsection{Network Learning}
Since implicit feedback in recommendation appears more common in real scenarios, we apply our model with BPR loss~\cite{rendle2009bpr} which is a powerful pairwise metric for implicit feedback data. And we adopt the same mini-batch negative sampling approach introduced in GRU4Rec. Suppose $u$ is one user, $p$ and $q$ denote the positive and negative item respectively. Therefore, the preference of user $u$ for positive item over negative item at time step $t$ is $\hat{r}_{upq}^t$, which is represented as:
\begin{equation}
\hat{r}_{upq}^t=(h_u^t)^T (z_{p}^{t+1} - z_{q}^{t+1}) 
\end{equation}
where $h_u^t$ is regarded as user representation at time step $t$, $z_{p}^{t+1}$ and $z_{q}^{t+1}$  are respectively the spectral graph embedding of positive sample item and negative sample item at $t+1$ step. The model aims to predict which item the user may interact with at next step, thus the supervised information comes from its next step.

The objective function is parameterized as $L_{CAASR}$:
\begin{equation}
L_{CAASR}=\sum_{(u,p,q)}-\ln(\sigma(\hat{r}_{upq}))+\frac{\lambda_{\Omega}}{2}||\Omega||_F^2
\label{eq:loss}
\end{equation}
where $\sigma(\hat{r}_{upq})$ indicates the probability that user $u$ prefers item $p$ than item $q$. $\Omega$ denotes the network parameters. $\frac{\lambda_{\Omega}}{2}||\Omega||_F^2$ plays as the regularization term in this objective function. $\lambda_{\Omega}\geq 0$ is the hyperparameter for regularization. A concise description of CAASR is illustrated in Algorithm~\ref{alg:CAASR}
 \begin{algorithm}[t]
 	\DontPrintSemicolon
 	\KwIn{1) An item adjacency matrix $A\in R^{N \times N}$ constructed from user behavior; \\
 	      2) An item feature matrix $X\in R^{N \times F}$, $X$ could be the identity matrix $I_{N}$ if there is no feature, and we adopt $X=I_{N}$ in all our experiments; \\
 	      3) Kernel size $K$ for convolution operation; \\
 	      4) Training sequence data from user behavior.}
 	\KwOut{For each interaction at step $t$ of a sequence, give the top $M$ recommended items that the user most probably might interact with at next step $t+1$.}
 	Calculate the Chebshev polynomial $T_{K}(\widetilde{L})$ by Eq.~\ref{equation:chebyshev polynomials} to support convolution operation.  \\
	\While{not converged}
	{   \For {$i$ in \{1,2,...,max-batch\}}
	    {   \textit{\# training triplets} \\
	        Generate training triplets $(u,p,q)$ from batch sequence data, where $r_{up}=1,r_{uq}=0$ and $r_{uq}=0$ is sampled through mini-batch sampling method depicted in GRU4Rec. \\
    	    \textit{\# association augmented graph embedding} \\
        	Calculate item association graph embedding $Z$ by Eq.~\ref{equation:X_convolution} ;  \\
        	Sample mini-batch item association embedding $z^{t}$ at step $t$ through embedding lookup layer depicted in Section~\ref{sec:emebdding_lookup_layer}; \\
        	\textit{\# general sequential recommendation model} \\
        	According to Section~\ref{sec:sequential_recommendation}, apply general RNNs based method (e.g. GRU4Rec) to obtain user representation $h_{u}^{t}$; \\ 
        	Update model parameters based on loss defined in Eq.~\ref{eq:loss}.
        }
    }
  \caption{CAASR}
 \label{alg:CAASR}
 \end{algorithm}

\subsection{Connection to GRU4Rec}
From Figure \ref{figure:model_architecture}, we can see that GRU4Rec is the sequential recommendation component of our model. In our method, the Chebyshev polynomial $T_k(x)$ degrades to $I_N$ when $K$=0. In this case, Eq.~\ref{equation:X_convolution} can be written as:
\begin{equation}
\begin{split}
Z=g_{\theta} * X \approx T_0(\tilde{L})X\Theta_{0} = X\Theta_{0},
\end{split}
\end{equation}
where $\Theta_0\in R^{C\times d}$ is exactly same to the embedding parameter for input $X$ in GRU4Rec. 
That is to say our CAASR can degrade to general RNNs based sequential recommendation when $K$=0. 

In addition, when the number of filters $K=0$, the spectral filter takes zero-hop of item's neighborhoods, meaning no consideration of the item's neighborhoods and item association information. Our model with a $K>0$ order Chebyshev polynomial can capture a high order association pattern among items.

\subsection{Parallel Style Association Augmented Models}
Apart from the proposed CAASR method in cascading style, item association relationships could also be combined with sequential relationships in two kinds of parallel styles. One typical way of incorporating it is co-factorization developed from non-sequential recommendation. And the other parallel style is the Autoencoder with $L_{2}$ regularization which is widely used in many deep learning models aiming to merge one specific information to the other. We would give a concise introduction about these two parallel styles to better demonstrate the advantages of our CAASR in cascading style.

\subsubsection{\textbf{Co-factorization Regularized}}
\begin{figure*}
    \centering
    \includegraphics[width=4.0in]{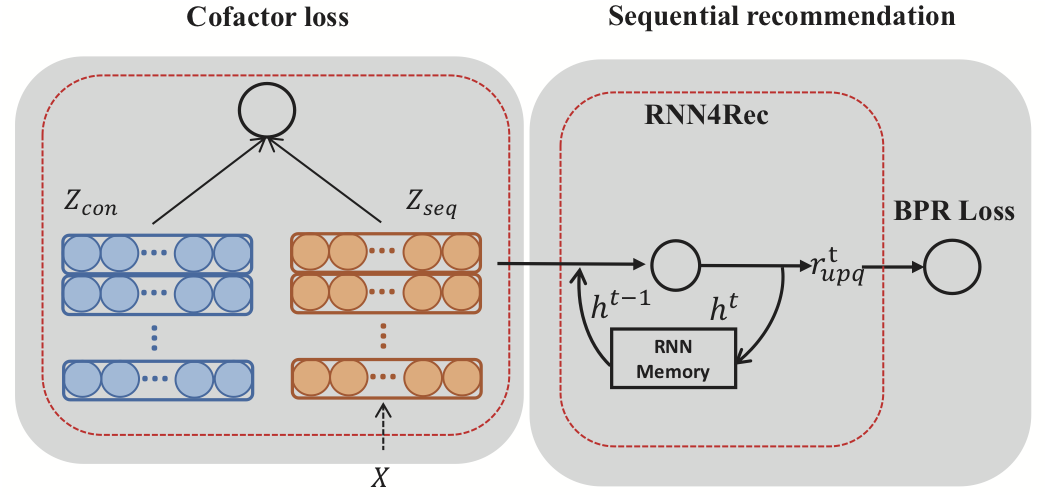}
    \caption{The architecture of our proposed P-Cofactor model. P-Cofactor directly introduces the co-factorization technique from non-sequential recommendation to mine item association relationships and combines it with sequential relationships in parallel style.}
    \label{figure:P_Cofactor}
\end{figure*}

Inspired by the widely used co-factorization technique in non-sequential recommendation, we transfer it to sequential recommendation to introduce the item association information. This combination model named P-Cofactor is depicted in Figure~\ref{figure:P_Cofactor}. 

Following method in~\cite{liang2016factorization}, the point-wise mutual information (\textit{PMI}) between item $\mathcal{I}_{i}$ and $\mathcal{I}_{j}$ is defined as:
\begin{equation}
    PMI(\mathcal{I}_{i}, \mathcal{I}_{j})=log\frac{P(\mathcal{I}_{i},\mathcal{I}_{j})}{P(\mathcal{I}_{i})P(\mathcal{I}_{j})}
\end{equation}
Empirically, it is estimated as:
\begin{equation}
    PMI(\mathcal{I}_{i},\mathcal{I}_{j})=log\frac{\#(\mathcal{I}_{i},\mathcal{I}_{j})\cdot \#pairs}{\#(\mathcal{I}_{i})\cdot \#(\mathcal{I}_{j})}
\end{equation}
where $\#(\mathcal{I}_{i},\mathcal{I}_{j})$ denotes the number of times that item $\mathcal{I}_{i}$ and $\mathcal{I}_{j}$ co-occur. $\#(\mathcal{I}_{i})=\sum_{j}\#(\mathcal{I}_{i},\mathcal{I}_{j})$ and $\#(\mathcal{I}_{j})=\sum_{i}\#(\mathcal{I}_{i},\mathcal{I}_{j})$. $\#pairs$ is the total number of co-occur pairs. Then the (sparse) Shifted Positive Point-wise Mutual Information (SPPMI) is defined as:
\begin{equation}
SPPMI(\mathcal{I}_{i},\mathcal{I}_{j})=max\{PMI(\mathcal{I}_{i},\mathcal{I}_{j})-log(o), 0\}
\label{eq:SPPMI}
\end{equation}
where $o$ is the hyperparameter that controls the sparsity of \textit{SPPMI} matrix. If we denote the item \textit{SPPMI} matrix as $S$, the loss function $L_{P-Cofactor}$ for P-Cofactor could be formalized as Eq.~\ref{eq:loss_P_Cofactor}.
\begin{equation}
    L_{P-Cofactor}=L_{s}+\underbrace{\sum_{s_{ij\neq 0}}(S_{ij}-Z_{con,i}\cdot Z_{seq,j}^{T})^{2}}_{SPPMI\ matrix\ factorization}
\label{eq:loss_P_Cofactor}
\end{equation}
where $L_{s}$ denotes the typical sequential recommendation module loss sharing the same formula in Eq.~\ref{eq:loss}. And $Z_{seq}\in R^{N\times d}$ is the item embedding shared by both the sequential recommendation module and \textit{SPPMI} matrix factorization module. $Z_{con}\in R^{N\times d}$ is the item context embedding for $Z_{seq}$. From the definition of $L_{P-Cofactor}$, we could summarize that this co-factorization technique is actually performing regularization for the learning process of item embedding in sequential recommendation. The concise learning process is illustrated in Algorithm~\ref{alg:P-Cofactor}.
 \begin{algorithm}[t]
 	\DontPrintSemicolon
 	\KwIn{1) The \textit{SPPMI} matrix $S\in R^{N \times N}$ defined in Eq.~\ref{eq:SPPMI}; \\
 	    2) Training sequence data from user behavior.}
 	\KwOut{For each interaction at step $t$ of a sequence, give the top $M$ recommended items that the user most probably might interact with at next step $t+1$.}
	\While{not converged}
	{   \For {$i$ in \{1,2,...,max-batch\}}
	    {   \textit{\# training triplets} \\
	        Generate training triplets $(u,p,q)$ input sequence behavior, where $r_{up}=1,r_{uq}=0$ and $r_{uq}=0$ is sampled through mini-batch sampling method depicted in GRU4Rec. \\
    	    \textit{\# parallel co-factorization model} \\
        	Update model parameters based on loss $L_{P-Cofactor}$ defined in Eq.~\ref{eq:loss_P_Cofactor}.
        }
    }
  \caption{P-Cofactor}
 \label{alg:P-Cofactor}
 \end{algorithm}
 
\subsubsection{\textbf{Graph Autoencoder Regularized}}
\begin{figure*}
    \centering
    \includegraphics[width=5.0in]{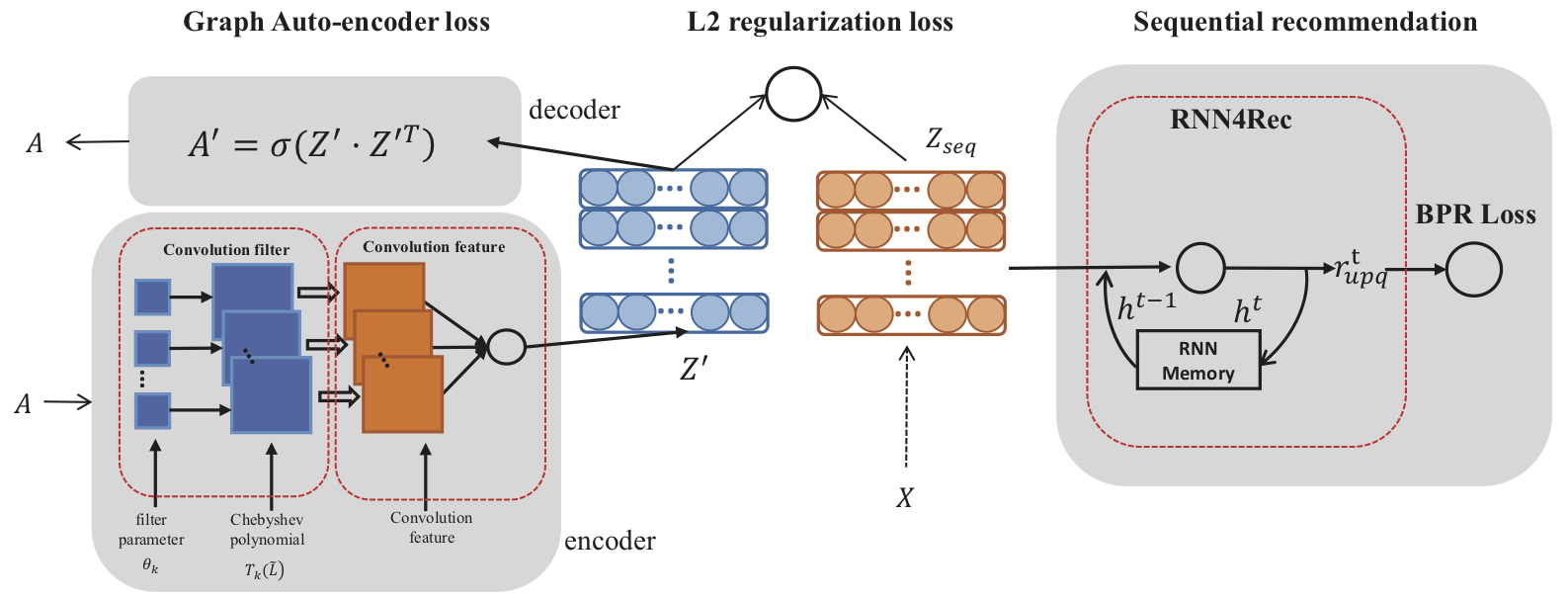}
    \caption{The architecture of our proposed P-GraphAE model. P-GraphAE utilizes graph Autoencoder and popular $L_{2}$ regularization technique to merge item association relationships and sequential relationships in parallel style.}
    \label{figure:P_GraphAE}
\end{figure*}

As one kind of unsupervised learning methods, Autoencoder, together with $L_{2}$ regularization, has been widely applied in combining two kinds of knowledge, such as Stacked Denoising Autoencoder (SDAE)~\cite{wang2015collaborative} and Variational Autoencoder (VAE)~\cite{li2017collaborative} in recommendation. Inspired by this, we construct item graph Autoencoder here to extract item association pattern in low-dimensional vector representation. Together with $L_{2}$ regularization technique, the architecture of graph Autoencoder regularized model named P-GraphAE is shown in Figure~\ref{figure:P_GraphAE}.

In P-GraphAE, Graph Autoencoder (GraphAE) is an essential part which mines the inherent association relationships in item graph. And the graph convolution operation in encoder is exactly the same with description in Section~\ref{sec:ASSR}. To make a clear description, we denote $Z^{'}$ as the latent representation of nodes in item graph for P-GraphAE and it shares the same definition in  Eq.~\ref{equation:X_convolution} as:
\begin{equation}
    Z^{'}=\sum_{k=0}^{K}T_{k}(\tilde{L})X\Theta_k^{'}
\label{eq:grah_encoder}
\end{equation}

As for the decoder, the aim lies in reconstructing the links in item graph. In practice, reconstructing all non-existent links in $A$ would easily lead to over-fitting, thus we randomly sample $m$ non-existent links to reconstruct. In our experiments, $m$ is 5 times larger than the number of existent links for all three datasets. Denote the training links set as $\mathcal{C}$, we could formulate the loss of P-GraphAE as:
\begin{equation}
\begin{split}
    L_{P-GraphAE}=& L_{s} + \underbrace{\sum_{(i,j)\in \mathcal{C} }-[A_{ij}log(A^{'}_{ij})+(1-A_{ij})log(1-A_{ij}^{'})]}_{GraphAE\ loss} \\ 
    & + \underbrace{||Z^{'}-Z_{seq}||_{F}^{2}}_{embedding\ regularization}
\end{split}
\label{eq:P_GraphAE_loss}
\end{equation}
where $L_{s}$ denotes the typical sequential recommendation module loss sharing the same formula in Eq.~\ref{eq:loss}. And $A^{'}=sigmoid(Z^{'}\cdot {Z^{'}}^{T})$ is the predicted adjacent matrix. $Z^{'}$ and $Z_{seq}$ denotes the latent item representation for GraphAE and sequential recommendation module respectively. $||Z^{'}-Z_{seq}||_{F}^{2}$ performs $L_{2}$ regularization and merges the item association information from item graph to sequential recommendation module. A concise illustration of this P-GraphAE is shown in Algorithm~\ref{alg:P-GraphAE}.
 \begin{algorithm}[t]
 	\DontPrintSemicolon
 	\KwIn{1) An item adjacency matrix $A\in R^{N \times N}$ constructed from user behavior; \\
 	      2) An item feature matrix $X\in R^{N \times F}$, $X$ could be the identity matrix $I_{N}$ if there is no feature, and we adopt $X=I_{N}$ in all our experiments; \\
 	      3) Kernel size $K$ for convolution operation; \\
 	      4) Training sequence data from user behavior.}
 	\KwOut{For each interaction at step $t$ of a sequence, give the top $M$ recommended items that the user most probably might interact with at next step $t+1$.}
 	{Calculate the Chebshev polynomial $T_{K}(\widetilde{L})$ by Eq.~\ref{equation:chebyshev polynomials} to support convolution operation.  }\\
	\While{not converged}
	{   \For {$i$ in \{1,2,...,max-batch\}}
	    {   \textit{\# training triplets} \\
	        Generate training triplets $(u,p,q)$ from batch sequence data, where $r_{up}=1,r_{uq}=0$ and $r_{uq}=0$ is sampled through mini-batch sampling method depicted in GRU4Rec. \\
    	    \textit{\# parallel Graph Autoencoder model} \\
        	Calculate item association graph embedding $Z^{'}$ by Eq.~\ref{eq:grah_encoder} in encoder;  \\
        	Update model parameters based on loss defined in Eq.~\ref{eq:P_GraphAE_loss}.
        }
    }
  \caption{P-GraphAE}\label{alg:P-GraphAE}
 \end{algorithm}

\subsection{Comparison Between Cascading Style and Parallel Style}
The way of combining this item association relationships and sequential relationships matters a lot both on model performance and complexity. Recent typical way of introducing associated information to existed recommendation involves parallel regularization of the corresponding representation~\cite{liang2016factorization, cao2017embedding}. This parallel regularization style puts a distance restriction on two different representations from distinctive domains. It is impractical to choose one specific distance manner and tune the regularization hyperparameter. Instead, the item association relationships describe the relational features for each item, which indicates the inherent knowledge of items. Hence, a cascading way to combine this item association relationships with sequential modeling is expected to perform better. In addition, parallel regularization involves additional reconstruction and regularization operation of the item graph, which makes it gain more complexity than cascading style.

\section{Experiments and Results}
\subsection{Datasets and Experimental Settings}
\subsubsection{\textbf{Datasets}}
To demonstrate the effectiveness of our model, we use three real-world datasets: Amazon Instant Video (AIV)\footnote{http://jmcauley.ucsd.edu/data/amazon/links.html\label{amazon_dataset}}, Amazon Cell Phones and Accessories (ACPA)\textsuperscript{\ref{amazon_dataset}} and TaoBao\footnote{https://tianchi.aliyun.com/datalab/dataSet.html?spm=5176.100073.0.0.61c835eeI6T5UL\&dataId=52}.

AIV and ACPA are both collected by MCAuley et.al~\cite{he2016ups} and these datasets contain explicit feedbacks with ratings range from 1 to 5 from Amazon\footnote{https://www.amazon.com/} during 1996.05 and 2014.07. In our experimental settings, we binary all the explicit feedbacks.

Taobao is a dataset from clothing matching competition on TianChi\footnote{https://tianchi.aliyun.com/} platform. The user purchase history, from 2014.6.14 to 2015.6.15, is considered as our sequence data here. Both these three datasets contain URLs to images of products. Note that in our experiments, we utilize these images for analysis rather than for training. We only use user-item interactions with timestamps to train our model.

All these three datasets need to be preprocessed for better model learning. Following~\cite{Hidasi2015Session}, we filtered users less than $n$ feedbacks and items less than $m$ interactions for all datasets. $n$ is 5,15,20 for AIV, ACPA and Taobao and $m$ is 5,10,10 for AIV, ACPA and Taobao respectively. Especially for TaoBao dataset, we find a serious click farming problem on it. Thus we also filter users that $n>75$. Users whose number of unique items in history less than 10 are also filtered.
\begin{table}[h!]
\caption{Statistics of three datasets after preprocessing. Interactions are non-zero entries. Data density and graph density refer to the density of user-item interaction matrix and adjacent matrix of item co-occurrence graph.}
\label{table:datasets_statistics}
\centering
\begin{tabular}{@{}cccc@{}}
\toprule
 & AIV & ACPA & TaoBao \\ \midrule
\#users & 6,798 & 1,273 & 17,400 \\
\#items & 4,733 & 8,816 & 15,366 \\
\#interactions & 50,858 & 25,427 & 477,447 \\
avg.len & 7.517 & 20.911 & 27.451 \\
data density & 0.158\% & 0.226\% & 0.178\% \\
graph density & 0.025\% & 0.011\% & 0.034\% \\ \bottomrule
\end{tabular}
\end{table}

After preprocessing, the statistics of these three datasets are summarized in Table~\ref{table:datasets_statistics}. Following~\cite{Hidasi2015Session}, for each dataset, we randomly sample 80\% sequences as train data, and the rest 20\% as test data. Note that one sequence is either in the train set or the test set. As it is a common operation in recommendation research, items not seen during training stage are filtered out for the test set.

\subsubsection{\textbf{Evaluation Metrics}}
Sequential Recommendation is usually with implicit feedback, in which we are expected to correctly predict the next item that the user will probably interact with. A good recommendation means that the target item should be among the first few recommended items. Thus in accordance with recent recommendation evaluation metrics, we adopt Recall@k and MRR@k as our evaluation metrics.

\subsubsection{\textbf{Baselines}}
We compare our CAASR with the following baselines:
\begin{itemize}
    \item {\bf BPR.} Bayesian Personalized Ranking (BPR)~\cite{rendle2009bpr} is one matrix factorization method for implicit feedback. BPR cannot be directly applied to sequential recommendation because new user does not have his or her representation vector. Thus following by GRU4Rec, we regard the average item feature vectors of items in user's history as the user representation.
    \item{\bf BPR+KNN.} K-Nearest Neighborhoods (KNN) is one common method in many practical Recommendation Systems. By learning item vectors in BPR model, we can utilize KNN at each time step to predict what the next item that the user will interact with. The similarity among items is defined as cosine similarity. This provides a baseline that concentrates on user's short-term interest compared to BPR.
    \item{\bf GRU4Rec.} GRU4Rec~\cite{Hidasi2015Session} is one typical RNNs based sequential recommendation method which captures the sequential relationships in user's sequential data. It also supports also supports user cold-start problem due to the fact that it does not specify a user in training process, which is practical in real scenarios. Outperformed performance makes it become one state-of-the-art RNNs based method for sequential recommendation.  
    \item{\bf P-Cofator.} Following item association relationships mining approach in non-sequential recommendation~\cite{liang2016factorization}, P-Cofactor introduces co-factorization technique to incorporate item association relationships and sequential relationships in sequential recommendation. This co-factorization technique actually performs one kind of regularization for RNNs based sequential recommendation.
    \item{\bf P-GraphAE.} P-GraphAE is the parallel extension of our cascading CAASR model, which applies popular Autoencoder and $L_{2}$ regularization approach to combine item association relationships and sequential relationships in sequential recommendation. Apart from P-Cofactor, it serves as the other parallel style to make comparison with CAASR.
\end{itemize}
Remind that our CAASR concentrates on modeling both item association relationships and sequential relationships in sequential user behavior, and we adopt GRU4Rec as the sequential relationships modeling method. This makes GRU4Rec become a perfect contrast baseline for CAASR. P-Cofactor and P-GraphAE are methods aiming to incorporate item association relationships with sequential relationships parallelly. Thus they form as the parallel baseline methods for our cascading method.

\subsubsection{\textbf{Experimental Settings}}
For all models, we set the maximum iteration up to 30. Batch training size is 50. Latent dimension size $d$ ranges in [50,100,150,200,
250,300], and the best $d$ is chosen for each model according to the performance on test set.

For BPR and BPR+KNN, we use a $L_2$ regularization term and the hyper-parameter is 0.01 to avoid over-fitting. Only one GRU layer is utilized for GRU4Rec and models based on GRU4Rec. For simplicity, we set $\lambda_{\Omega}=0$ and apply Dropout~\cite{srivastava2014dropout} technique for regularization. Dropout rate is 0.2 for all deep learning based methods. Parameters are initialized with uniform distribution ranges from -0.1 to 0.1. RMSprop is adopted for optimization. Table~\ref{table:Hyperparameters_setting} shows the hyper-parameter setting for all models on three datasets.
\begin{table}
\caption{Hyper-parameter setting for all models used in our experiments on three datasets. The Chebyshev polynomial order $K$ for our model ranges in [3,4,5]. Limited to our computational resource, $K=3$ is adopted for TaoBao without testing $K=4,5$. Dropout rate means the value of drop rate.}
\label{table:Hyperparameters_setting}
\centering
\begin{tabular}{llll}
\hline
\textbf{Parameter}                 & \textbf{AIV} & \textbf{ACPA} & \textbf{TaoBao} \\ \hline
\textbf{BPR/BPR+KNN}               &              &               &                 \\
latent factors                     & 50/300       & 50/250        & 300/300         \\
learning rate                      & 0.01/0.01    & 0.01/0.01     & 0.01/0.01       \\
regularization                     & 0.01/0.01    & 0.01/0.01     & 0.01/0.01       \\
\textbf{GRU4Rec/P-Cofactor} &              &               &                 \\
hidden units                       & 50/50       & 250/50       & 150/100         \\
learning rate                      & 0.01/0.01   & 0.01/0.01    & 0.01/0.01      \\
Dropout rate                            & 0.2/0.2      & 0.2/0.2       & 0.2/0.2         \\
\textbf{P-GraphAE/CAASR} &              &               &                 \\
hidden units                       & 100/300       & 150/300       & 300/300         \\
learning rate                      & 0.001/0.001   & 0.001/0.001    & 0.001/0.001      \\
Dropout rate                            & 0.2/0.2      & 0.2/0.2       & 0.2/0.2         \\
K-order                            & 5/5          & 4/4           & 3/3 \\
\hline
\end{tabular}
\end{table}

\subsection{Quantitative Results}
\begin{table*}
\caption{Overall performance of all models on three datasets. The relative improvement of our CAASR over GRU4Rec is given in the table. Model names with * refer to our proposed methods.  Compared to CAASR, the t-test results of other baselines are as well shown in the table. And $\ddagger$ means p-value<0.01, $\dagger$ indicates p-value<0.05 and $-$ means p-value>0.05.}
\label{table:overall_performance}
\centering
\begin{tabular}{|c|c|c|c|c|c|}
\hline
\multirow{2}{*}{Dataset} & \multirow{2}{*}{Method} & \multicolumn{2}{c|}{@10}                                  & \multicolumn{2}{c|}{@20}                                  \\ \cline{3-6} 
                         &                         & Recall                      & MRR                         & Recall                      & MRR                         \\ \hline
\multirow{4}{*}{AIV}     & BRR                     & 0.0643$^\ddagger$                    & 0.0189$^\ddagger$                    & 0.0793$^\ddagger$                    & 0.0200$^\ddagger$                    \\ \cline{2-6} 
                         & BPR+KNN                 & 0.1041$^\ddagger$                    & 0.0314$^\ddagger$                    & 0.1280$^\ddagger$                    & 0.0331$^\ddagger$                    \\ \cline{2-6} 
                         & GRU4Rec                 & 0.1084$^\ddagger$                    & 0.0505$^\ddagger$                    & 0.1505$^\ddagger$                    & 0.0536$^\ddagger$                   \\ \cline{2-6}
                         & P-Cofactor$^{*}$                 & 0.1081$^\ddagger$                    & 0.0491$^\ddagger$                    & 0.1534$^\ddagger$                    & 0.0525$^\ddagger$                   \\ \cline{2-6}
                         & P-GraphAE$^{*}$                 & 0.1157$^\ddagger$                    & 0.0524$^\ddagger$                    & 0.1632$^\ddagger$                    & 0.0558$^\ddagger$                   \\ \cline{2-6}
                         & \textbf{CAASR}$^{*}$          & \textbf{\tabincell{c}{0.1271\\(+17.25\%)}} & \textbf{\tabincell{c}{0.0585\\(+15.84\%)}} & \textbf{\tabincell{c}{0.1778\\(+18.13\%)}} & \textbf{\tabincell{c}{0.0623\\(+16.23\%)}} \\ \hline
\multirow{4}{*}{ACPA}    & BRR                     & 0.0027$^\ddagger$                    & 0.0006$^\ddagger$                    & 0.0072$^\ddagger$                   & 0.0009$^\ddagger$                    \\ \cline{2-6} 
                         & BPR+KNN                 & 0.0303$^\ddagger$                    & 0.0089$^\ddagger$                    & 0.0423$^\ddagger$                    & 0.0098$^\ddagger$                    \\ \cline{2-6} 
                         & GRU4Rec                 & 0.0714$^\ddagger$                    & 0.0337$^\ddagger$                    & 0.0997$^\ddagger$                    & 0.0358$^\ddagger$                    \\ \cline{2-6}
                         & P-Cofactor$^{*}$                 & 0.0571$^\ddagger$                    & 0.0223$^\ddagger$                    & 0.0882$^\ddagger$                    & 0.0237$^\ddagger$                   \\ \cline{2-6}
                         & P-GraphAE$^{*}$                 & 0.0586$^\ddagger$                    & 0.0243$^\ddagger$                    & 0.0877$^\ddagger$                    & 0.0364$^\ddagger$                   \\ \cline{2-6}
                         & \textbf{CAASR}$^{*}$          & \textbf{\tabincell{c}{0.0827\\(+15.82\%)}} & \textbf{\tabincell{c}{0.0357\\(+5.93\%)}}  & \textbf{\tabincell{c}{0.1228\\(+23.16\%)}} & \textbf{\tabincell{c}{0.0387\\(+8.10\%)}}  \\ \hline
\multirow{4}{*}{TaoBao}  & BRR                     & 0.0683$^\ddagger$                    & 0.0250$^\ddagger$                    & 0.1004$^\ddagger$                    & 0.0272$^\ddagger$                    \\ \cline{2-6} 
                         & BPR+KNN                 & 0.0601$^\ddagger$                    & 0.0179$^\ddagger$                    & 0.0873$^\ddagger$                    & 0.0197$^\ddagger$                    \\ \cline{2-6} 
                         & GRU4Rec                 & 0.1472$^\ddagger$                    & 0.0721$^\ddagger$                    & 0.1984$^\ddagger$                    & 0.0759$^\ddagger$                    \\ \cline{2-6}
                         & P-Cofactor$^{*}$                 & 0.1406$^\ddagger$                    & 0.0691$^\ddagger$                    & 0.1897$^\ddagger$                    & 0.0727$^\ddagger$                   \\ \cline{2-6}
                         & P-GraphAE$^{*}$                 & 0.1578$^\ddagger$                    & 0.0700$^\ddagger$                    & 0.2109$^\ddagger$                    & 0.0746$^\dagger$                   \\ \cline{2-6}
                         & \textbf{CAASR}$^{*}$          & \textbf{\tabincell{c}{0.1654\\(+12.36\%)}} & \textbf{\tabincell{c}{0.0731\\(+1.38\%)}}  & \textbf{\tabincell{c}{0.2186\\(+10.18\%)}} & \textbf{\tabincell{c}{0.0771\\(+1.58\%)}}  \\ \hline
\end{tabular}
\end{table*}
\subsubsection{\textbf{Overall Performance}}
In this subsection, we show the performance of all models in Table~\ref{table:overall_performance}. To better highlight the performance of our model, the relative performance compared to GRU4Rec is also illustrated in Table~\ref{table:overall_performance}. 

(1) From this table, we can observe that deep learning based methods (GRU4Rec, P-GraphAE and CAASR) yield superior performance compared to BPR and BPR+KNN. It is reasonable due to the superior representation learning ability of deep learning.

(2) Our proposed model achieves the best performance on all three datasets. Compared to the state-of-the-art GRU4Rec, we even achieve a 23.16\% relative improvement at Recall@20 on ACPA and 16.23\% relative improvement at MRR@20 on AIV. Remind that the only difference between our CAASR and GRU4Rec is the additional item association embedding part. And this demonstrates the significance of modeling item association relationships for sequential user behavior.

(3) From the table, it is clear that P-Cofactor developed from GRU4Rec performs worse than GRU4Rec almost on each dataset. Especially on ACPA, it is even 1.43\% and 1.15\% lower than GRU4Rec on Recall@10 and Recall@20 respectively. Remind that P-Cofactor directly introduces co-factorization technique to incorporate the item association relationships into RNNs based sequential recommendation method. The result indicates that co-factorization technique from non-sequential recommendation actually harms the performance of GRU4Rec. This happens mainly because it is incompatible to apply $L_{2}$ regularization between deeply learned representation by RNNs and shallowly represented representation through traditional matrix factorization. 

(4) Compared to P-Cofactor, P-GraphAE and CAASR both adopt graph convolution operation. And the parallel style P-GraphAE generally performs slightly better than GRU4Rec on AIV and TaoBao, while it fails on ACPA. In contrast, our CAASR in cascading style always performs better than GRU4Rec and P-GraphAE on all datasets. Regarding Recall@20 on ACPA, it achieves a 2.31\% and 3.46\% improvement compared to GRU4Rec and P-GraphAE respectively. These results illustrate the correctness of cascading style rather than parallel style. In addition, compared to P-GraphAE, the cascading CAASR has lower complexity and generalizes the input of RNNs based sequential recommendation. This allows it to degrade to general RNNs based sequential recommendation model under specific condition while P-GraphAE fails.

(5) To verify whether the improvement of CAASR is statistical significant, we conduct \textit{t-test} here and the result of p-value in shown in Table.~\ref{table:overall_performance} with different markers $\ddagger$, $\dagger$ and $-$. In this table, p-value refers to the comparison between CASSR and other baselines. The p-values in Table.~\ref{table:overall_performance} are all less than 0.01 or 0.05, which validates the improvements of our CAASR are statistical significant.

\begin{figure*}[h!]
\centering
\begin{minipage}[t]{0.45\textwidth}
\centering
\includegraphics[width=\textwidth]{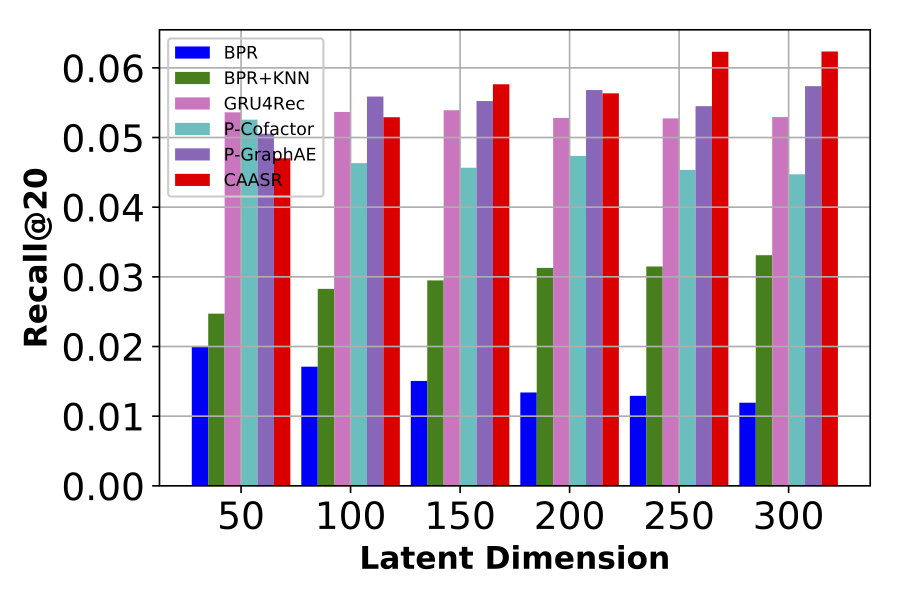}
\vspace{-20pt}
\caption*{(a) AIV - Recall@20}
\end{minipage}
\begin{minipage}[t]{0.45\textwidth}
\centering
\includegraphics[width=\textwidth]{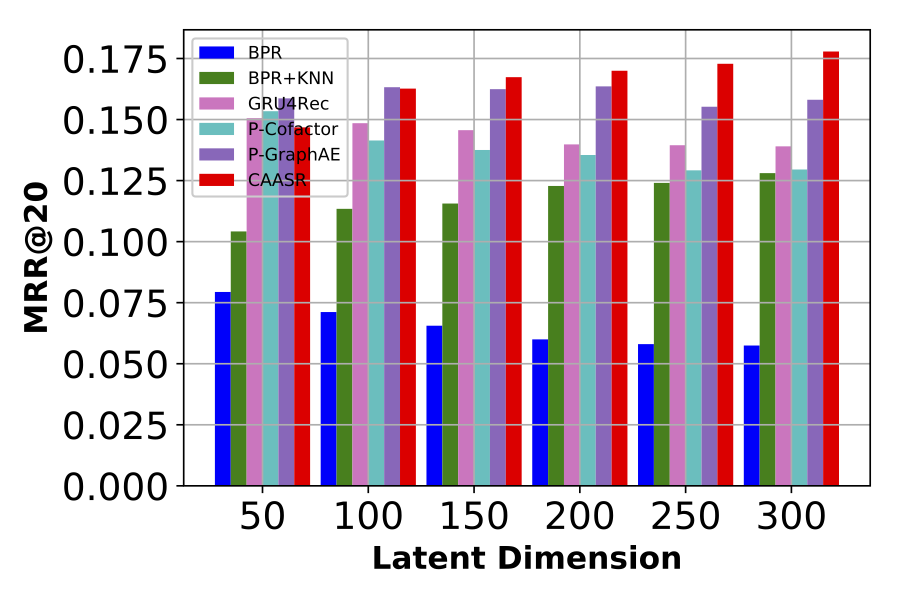}
\vspace{-20pt}
\caption*{(b) AIV - MRR@20}
\end{minipage} \\
\begin{minipage}[t]{0.45\textwidth}
\centering
\includegraphics[width=\textwidth]{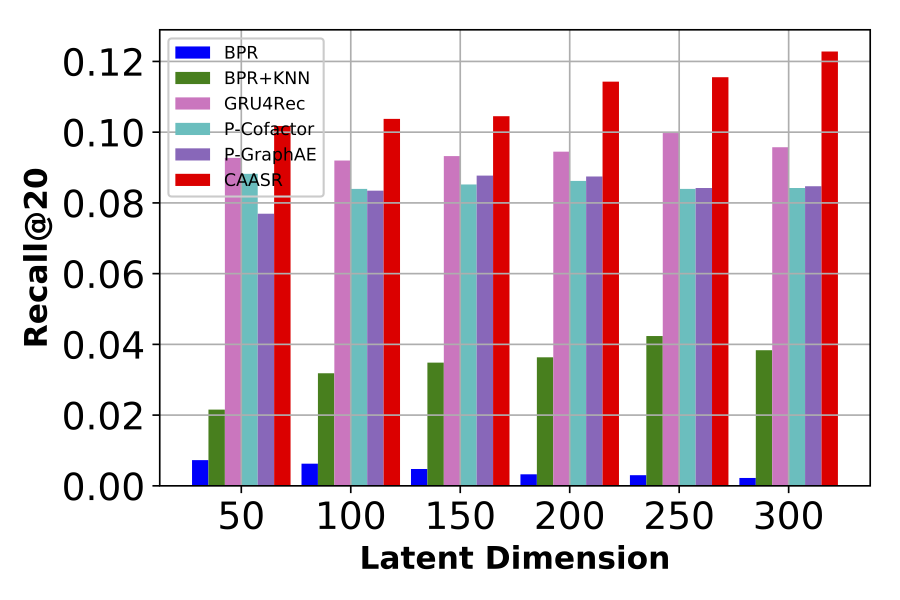}
\vspace{-20pt}
\caption*{(c) ACPA - Recall@20}
\end{minipage}
\begin{minipage}[t]{0.45\textwidth}
\centering
\includegraphics[width=\textwidth]{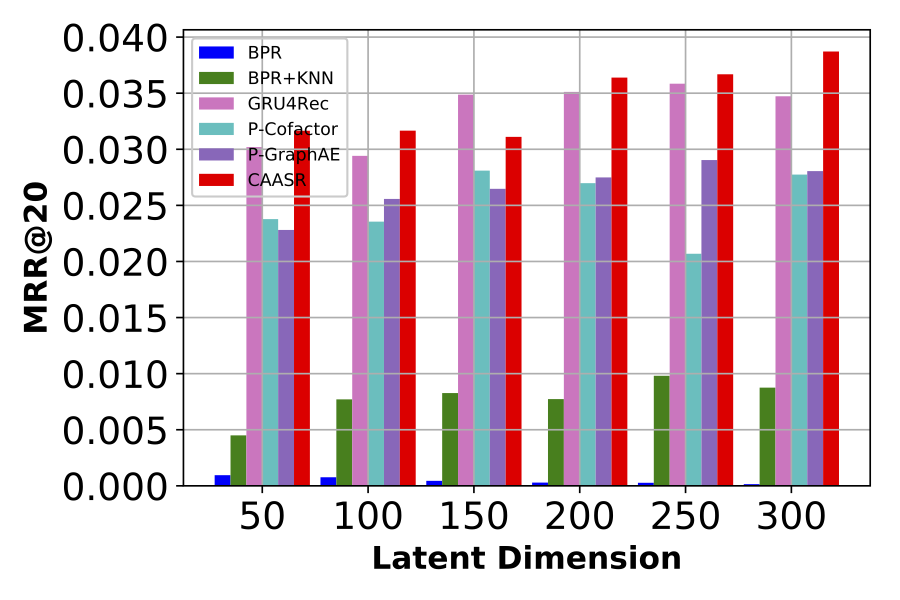}
\vspace{-20pt}
\caption*{(d) ACPA - MRR@20}
\end{minipage} \\
\begin{minipage}[t]{0.45\textwidth}
\centering
\includegraphics[width=\textwidth]{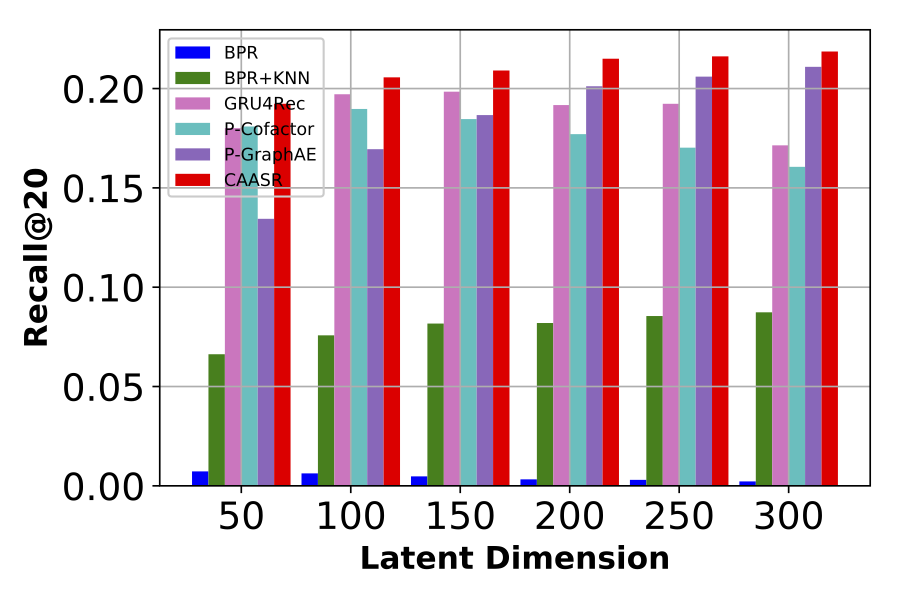}
\vspace{-20pt}
\caption*{(e) TaoBao - Recall@20}
\end{minipage}
\begin{minipage}[t]{0.45\textwidth}
\centering
\includegraphics[width=\textwidth]{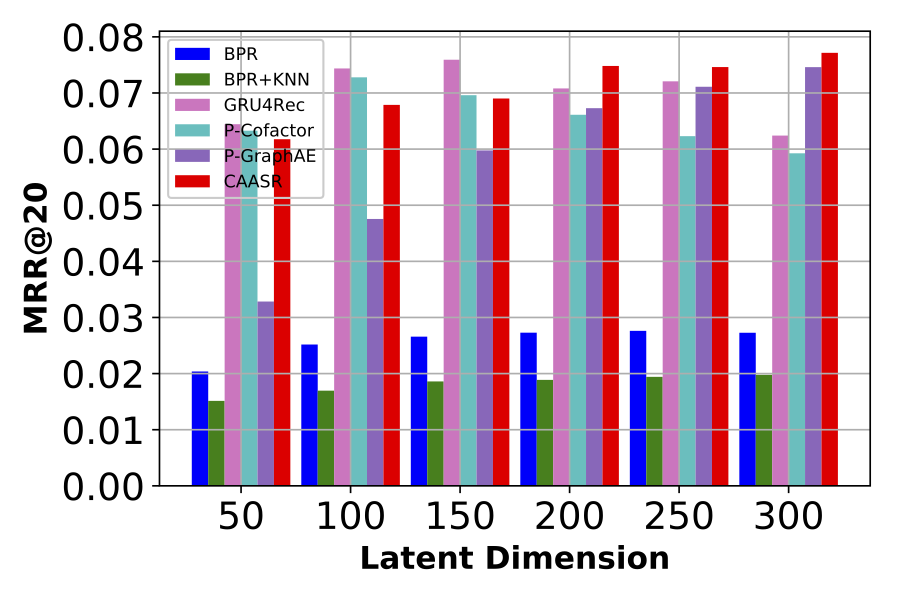}
\vspace{-20pt}
\caption*{(f) TaoBao - MRR@20}
\end{minipage}
\caption{Performance of Recall@20 and MRR@20 with different latent dimensions on three datasets. (a)(c)(e): Recall@20 on three datasets. (b)(d)(f): MRR@20 on three datasets.}
\label{figure:latent_dimension}
\end{figure*}

\subsubsection{\textbf{Performance on Different Latent Dimensions}}
In Figure~\ref{figure:latent_dimension}, for all models, we show their performance on Recall@20 and MRR@20 with different latent dimensions. In Figure~\ref{figure:latent_dimension} (a)(c)(e), we can see that our model reaches a better performance of Recall@20 on three datasets almost at every latent dimension. As for MRR@20 in Figure~\ref{figure:latent_dimension} (b)(d)(f), our model can still perform better when the latent dimension is high. Note that Recall@20 and MRR@20 are different kinds of evaluation metrics. Better Recall@20 does not mean an exactly better MRR@20. We compare our model both in recall ability and ranking ability.

Moreover, Figure~\ref{figure:latent_dimension} indicates that BRR, BPR+KNN and GRU4Rec model tend to have their best performance at low dimension. While our CAASR favors reaching its best performance at high dimension. This is probably because both BPR and GRU4Rec take deficient utilization of inputs. When the dimension is high, the information in learned vectors tends to be redundant and yields to be not robust on test set. While CAASR learns item association relationships in spectral graph domain with more information flow. This needs to match a higher dimension to adequately encode the data.

From Figure~\ref{figure:latent_dimension}, it is also obvious that P-Cofactor generally does worse performance than GRU4Rec at every dimension, enhancing the claim that co-factorization developed from non-sequential recommendation could not bring reliable gain for recommendation in RNNs based sequential recommendation. As for Recall@20 and MRR@20 of P-GraphAE, it shares a similar trend with CAASR which tends to achieve a better performance along with the increase of latent dimension. Whereas P-GraphAE with higher complexity still cannot perform better than CAASR, which emphasizes the advantage of the better cascading style.
% 指标随着epoch变化曲线图
\begin{figure*}[h!]
%第一行 loss
\centering
\begin{minipage}[t]{0.3\textwidth}
\centering
\includegraphics[width=\textwidth]{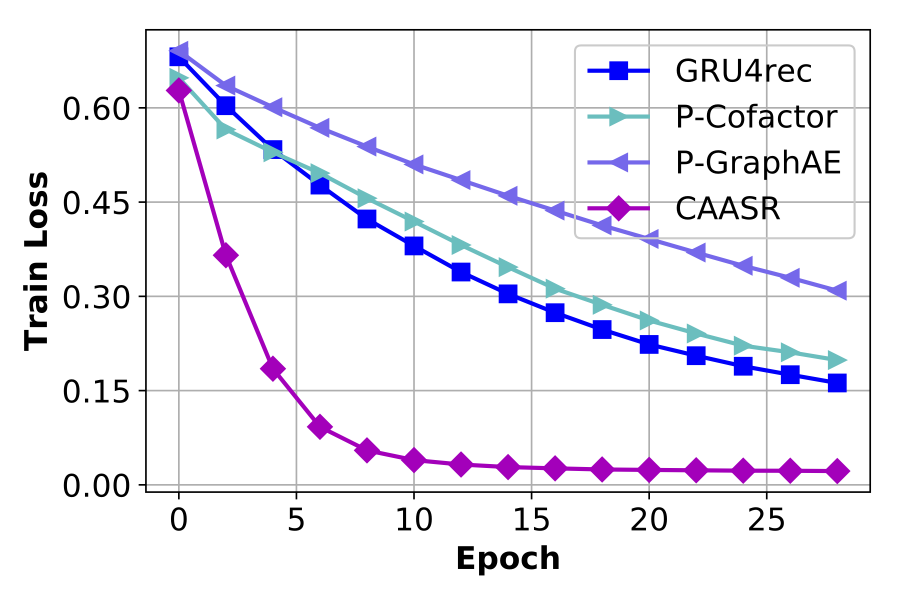}
\vspace{-22pt}
\caption*{(a) AIV - Train BPR Loss}
\end{minipage}
\begin{minipage}[t]{0.3\textwidth}
\centering
\includegraphics[width=\textwidth]{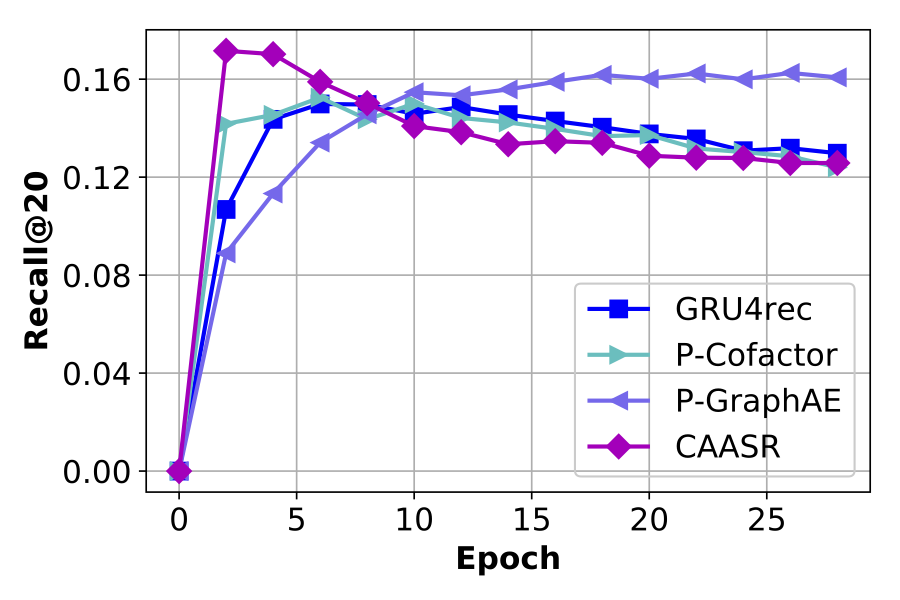}
\vspace{-22pt}
\caption*{(b) AIV - Recall@20}
\end{minipage}
\begin{minipage}[t]{0.3\textwidth}
\centering
\includegraphics[width=\textwidth]{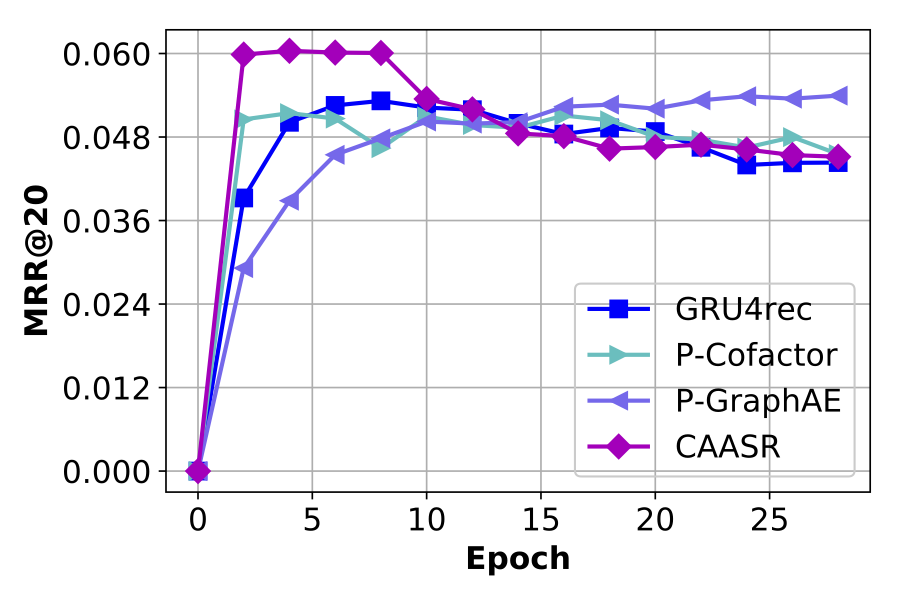}
\vspace{-22pt}
\caption*{(c) AIV - MRR@20}
\end{minipage}
%第二行 recall@20
\begin{minipage}[t]{0.3\textwidth}
\centering
\includegraphics[width=\textwidth]{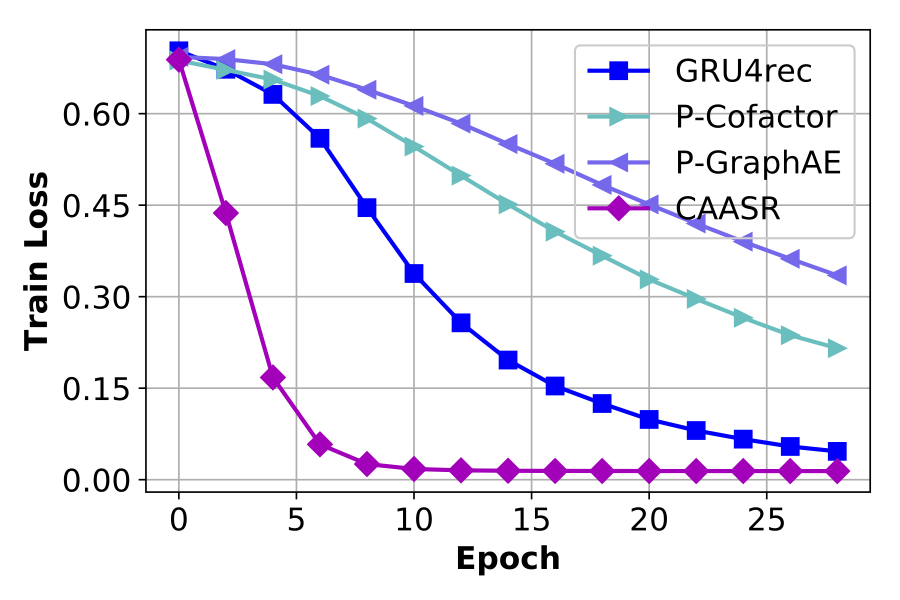}
\vspace{-22pt}
\caption*{(d) ACPA - Train BPR Loss}
\end{minipage}
\begin{minipage}[t]{0.3\textwidth}
\centering
\includegraphics[width=\textwidth]{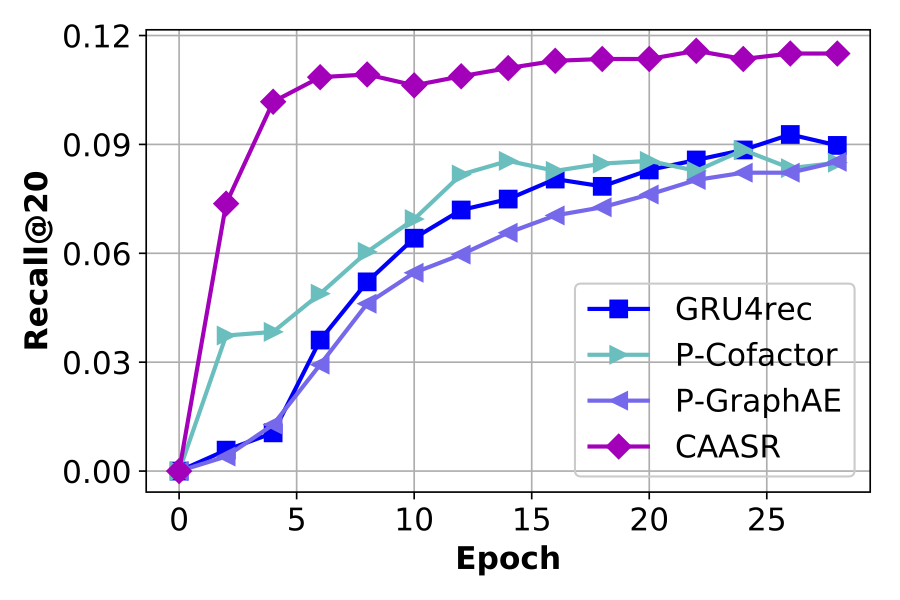}
\vspace{-22pt}
\caption*{(e) ACPA - Recall@20}
\end{minipage}
\begin{minipage}[t]{0.3\textwidth}
\centering
\includegraphics[width=\textwidth]{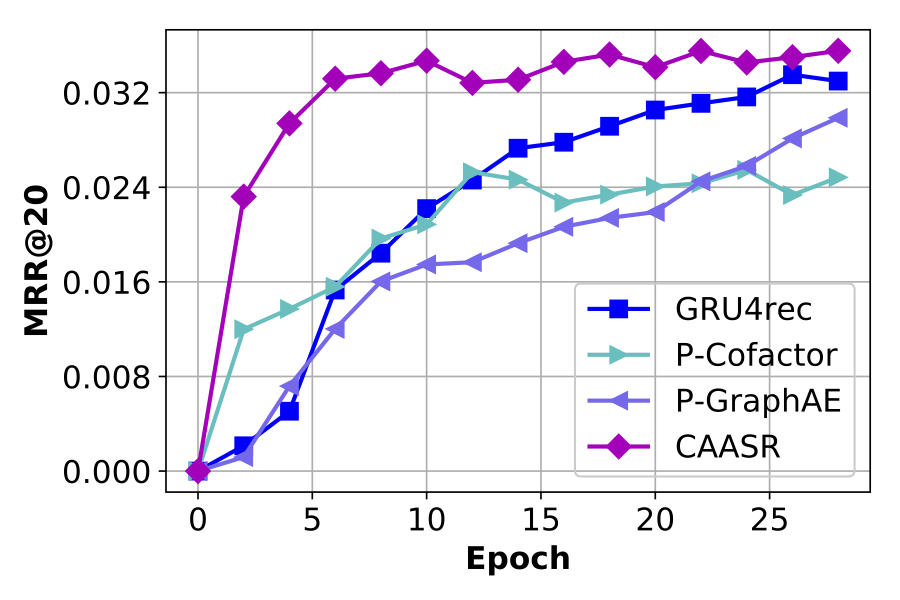}
\vspace{-22pt}
\caption*{(f) ACPA - MRR@20}
\end{minipage}
%第三行 mrr@20
\begin{minipage}[t]{0.3\textwidth}
\centering
\includegraphics[width=\textwidth]{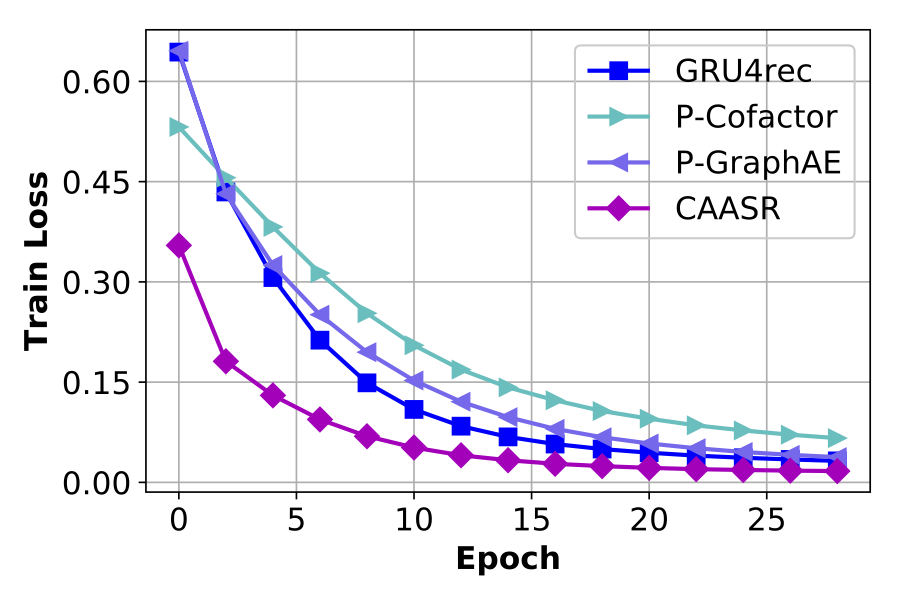}
\vspace{-22pt}
\caption*{(g) TaoBao - Train BPR Loss}
\end{minipage}
\begin{minipage}[t]{0.3\textwidth}
\centering
\includegraphics[width=\textwidth]{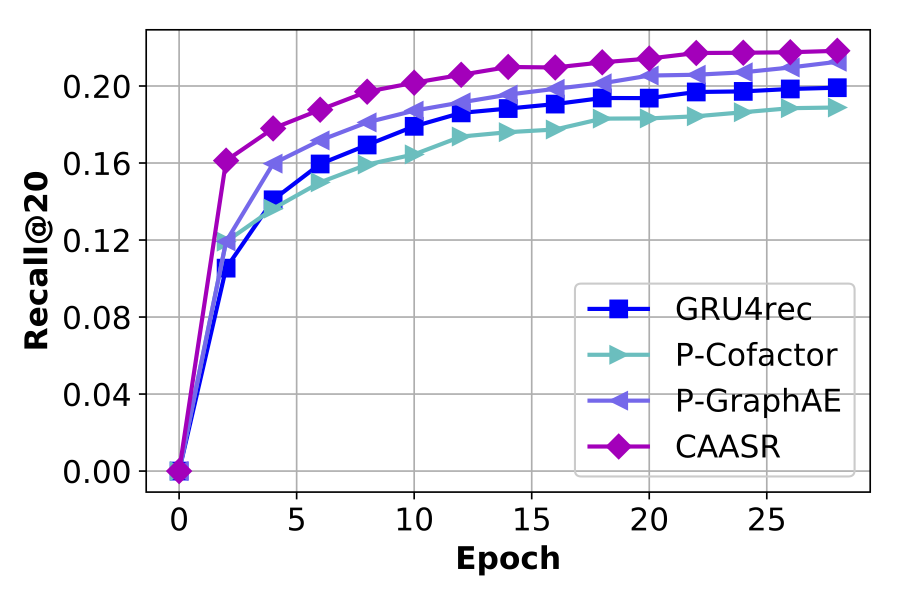}
\vspace{-22pt}
\caption*{(h) TaoBao - Recall@20}
\end{minipage}
\begin{minipage}[t]{0.3\textwidth}
\centering
\includegraphics[width=\textwidth]{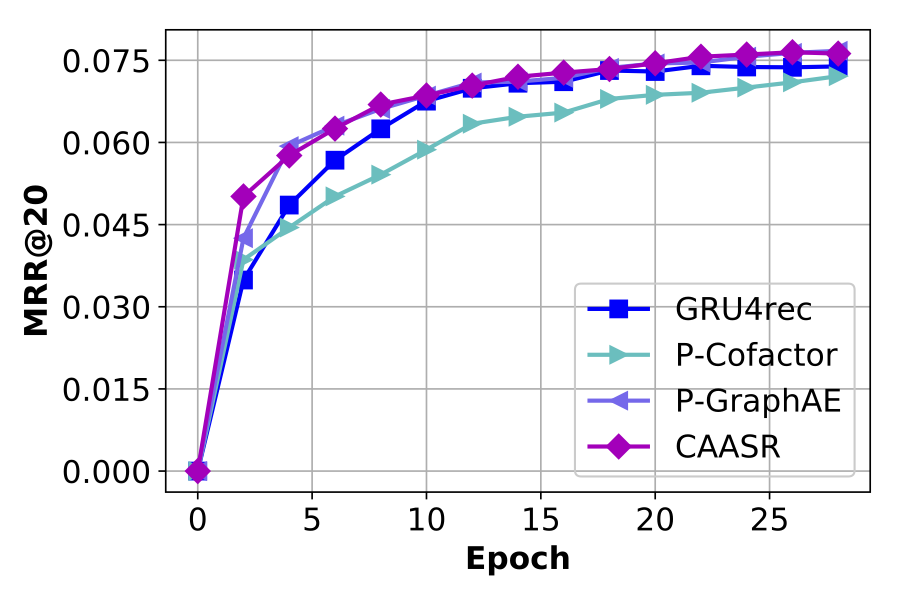}
\vspace{-22pt}
\caption*{(i) TaoBao - MRR\_20}
\end{minipage}
\caption{Train loss and performance of Recall@20 and MRR@20 with training epochs on three datasets. (a)(b)(c): train BPR loss on three datasets; (d)(e)(f): Recall@20 on three datasets; (g)(h)(i): MRR@20 on three datasets. The results of GRU4Rec, P-Cofactor, P-GraphAE and CAASR are reported here due to the fact that they share the same BPR loss in sequential training.}
\label{figure:performance_epochs}
\end{figure*}
\subsubsection{\textbf{Training Loss and Performance.}} In order to explore the influence of the association relationships and the way of combining it with sequential relationships captured by RNNs based sequential recommendation method, we show the train BPR loss, test Recall@20 and test MRR@20 along with training epochs on all three datasets in Figure~\ref{figure:performance_epochs}.

It is clear from Figure~\ref{figure:performance_epochs} (a)(d)(g) that CAASR converges faster and achieves a lower BPR loss on all three datasets than the other three methods. As for Recall@20  and MRR@20 on AIV shown in Figure~\ref{figure:performance_epochs} (b)(c), CAASR reaches its best performance at epoch less than 5 and later training causes over-fitting, while the other three models need more training epochs. As for performance on ACPA and TaoBao in Figure~\ref{figure:performance_epochs} (e)(f)(h)(i), CAASR outperforms the others along the training process and the gap is more obvious in less training epochs, which reflects the fast convergence rate of CAASR. In summary, combining item association relationships and sequential pattern in cascading style fastens the convergence rate of BPR loss, as well as helps the RNNs based sequential recommendation reach a better learning point for final recommendation target.

\begin{table}[!h]
\caption{Recall@20 of CAASR with different Chebshev order $K$ and latent dimension $d$ on two datasets.}
\label{table:Recall@20_K}
\centering
\begin{tabular}{|c|c|c|c|c|c|c|}
\hline
\textbf{Factors} & \textbf{50}    & \textbf{100}   & \textbf{150}   & \textbf{200}   & \textbf{250}   & \textbf{300}   \\ \hline
\multicolumn{7}{|c|}{\textbf{AIV}}                                                                                     \\ \hline
\textbf{K=3}     & 0.142          & 0.152          & 0.158          & 0.158          & 0.167          & 0.163          \\ \hline
\textbf{K=4}     & 0.142          & 0.161          & 0.166          & 0.167          & 0.167          & 0.174          \\ \hline
\textbf{K=5}     & \textbf{0.147} & \textbf{0.163} & \textbf{0.167} & \textbf{0.170} & \textbf{0.173} & \textbf{0.178} \\ \hline
\multicolumn{7}{|c|}{\textbf{ACPA}}                                                                                    \\ \hline
\textbf{K=3}     & 0.096          & 0.098          & 0.101          & 0.100          & 0.104          & 0.106          \\ \hline
\textbf{K=4}     & \textbf{0.102} & \textbf{0.104} & \textbf{0.105} & \textbf{0.114} & \textbf{0.116} & \textbf{0.123} \\ \hline
\textbf{K=5}     & 0.098          & 0.099          & 0.103          & 0.110          & 0.112          & 0.117          \\ \hline
\end{tabular}
\end{table}

\begin{table}[!h]
\caption{MRR@20 of CAASR with different Chebshev order $K$ and latent dimension $d$ on two datasets.}
\label{table:MRR@20_K}
\centering
\begin{tabular}{|c|c|c|c|c|c|c|}
\hline
\textbf{Factors} & \textbf{50}    & \textbf{100}   & \textbf{150}   & \textbf{200}   & \textbf{250}   & \textbf{300}   \\ \hline
\multicolumn{7}{|c|}{\textbf{AIV}}                                                                                     \\ \hline
\textbf{K=3}     & 0.045          & 0.048          & 0.050          & 0.054          & 0.056          & 0.055          \\ \hline
\textbf{K=4}     & \textbf{0.047} & 0.050          & 0.055          & 0.056          & 0.059          & 0.060          \\ \hline
\textbf{K=5}     & 0.046          & \textbf{0.053} & \textbf{0.058} & \textbf{0.056} & \textbf{0.062} & \textbf{0.062} \\ \hline
\multicolumn{7}{|c|}{\textbf{ACPA}}                                                                                    \\ \hline
\textbf{K=3}     & 0.027          & 0.030          & 0.029          & 0.031          & 0.033          & 0.029          \\ \hline
\textbf{K=4}     & \textbf{0.032} & \textbf{0.033} & \textbf{0.031} & \textbf{0.036} & \textbf{0.037} & \textbf{0.039} \\ \hline
\textbf{K=5}     & 0.031          & 0.032          & 0.031          & 0.033          & 0.032          & 0.035          \\ \hline
\end{tabular}
\end{table}
\subsubsection{\textbf{Parameter Analysis on Latent Dimension $d$ and Chebyshev Order $K$.}}
As the Chebyshev polynomial order $K$ is involved in our method, it is curious to see whether a higher Chebyshev polynomial order is beneficial to the recommendation accuracy. Towards this target, we further investigate our CAASR with different $K$ on two datasets: AIV and ACPA (TaoBao dataset is not implemented here due to our limited computational resource). The results of both Recall@20 and MRR@20 on two datasets are summarized in Table~\ref{table:Recall@20_K} and ~\ref{table:MRR@20_K}. 

As we can see when at the same model capability (with the same latent factors), with the increase of Chebyshev polynomial order $K$ ranges in [3,4,5], CAASR performs best at $K=5$ on AIV. While for ACPA, CAASR tends to reach its best performance at $K=4$. Remind that $K$ denotes the number of hops from the central node. Too large $K$ may lead to consider uninformative neighborhoods. Therefor the proper $K$ tends to be related to the specific dataset.

\begin{figure*}[h!]
\centering
\begin{minipage}[t]{0.32\textwidth}
\centering
\includegraphics[width=\textwidth]{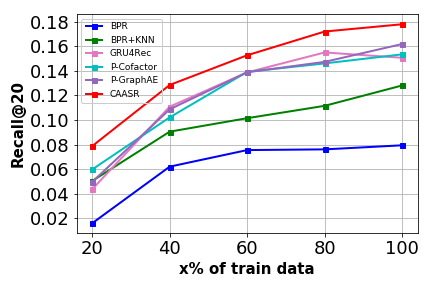}
\vspace{-20pt}
\caption*{(a) AIV - Recall@20}
\end{minipage}
\begin{minipage}[t]{0.32\textwidth}
\centering
\includegraphics[width=\textwidth]{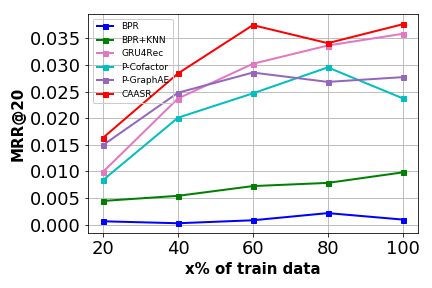}
\vspace{-20pt}
\caption*{(b) ACPA - Recall@20}
\end{minipage} 
\begin{minipage}[t]{0.32\textwidth}
\centering
\includegraphics[width=\textwidth]{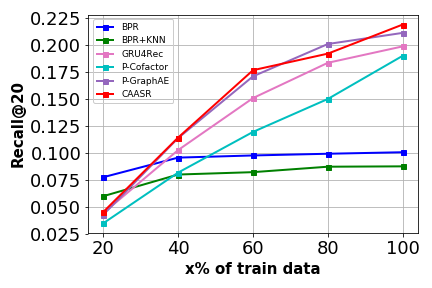}
\vspace{-20pt}
\caption*{(c) TaoBao - Recall@20}
\end{minipage} \\
\begin{minipage}[t]{0.32\textwidth}
\centering
\includegraphics[width=\textwidth]{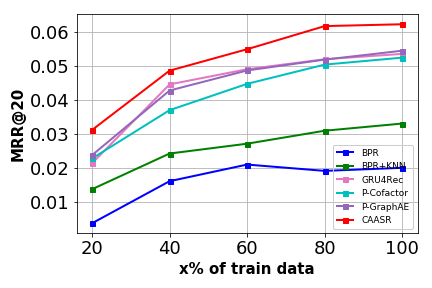}
\vspace{-20pt}
\caption*{(d) AIV - MRR@20}
\end{minipage}
\begin{minipage}[t]{0.32\textwidth}
\centering
\includegraphics[width=\textwidth]{train_ratio_ACPA_MRR_20.png}
\vspace{-20pt}
\caption*{(e) ACPA - MRR@20}
\end{minipage}
\begin{minipage}[t]{0.32\textwidth}
\centering
\includegraphics[width=\textwidth]{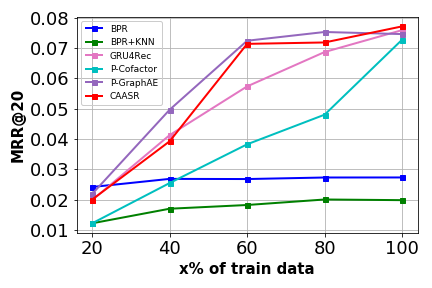}
\vspace{-20pt}
\caption*{(f) TaoBao - MRR@20}
\end{minipage}
\caption{Performance of Recall@20 and MRR@20 with different ratio of train data and Chebshev $K$ on three datasets. (a)(b)(c): Recall@20 on three datasets. (d)(e)(f): MRR@20 on three datasets.}
\label{figure:train_ratio}
\end{figure*}
\subsubsection{\textbf{Sparse Train Data.}}
Different sparse train data may have different influence on different methods. The sparse train data directly leads to sparse item graph, which makes it meaningful to explore whether our graph embedding based methods could still outperform general GRU4Rec. Therefore, we conduct an experiment about the model performance with different ratio of train data while keeping the test data fixed. Figure~\ref{figure:train_ratio} shows the corresponding result on AIV, ACPA and TaoBao.
 
 From Figure~\ref{figure:train_ratio}, we can see that, in general, all model performance rises with the increase of train data. Our cascading CAASR is capable of outperforming other methods under different train data conditions in terms of both general GRU4Rec method and the parallel methods (P-Cofactor and P-GraphAE). In Figure~\ref{figure:train_ratio} (c)(f) on TaoBao, P-GraphAE poses slightly better than CAASR under some train data ratio conditions while it certainly fails on AIV and ACPA in Figure~\ref{figure:train_ratio}(a)(b)(d)(e). Instead, the casacading shares a lower computational complexity and more robust performance than P-GraphAE, which verifies the meaning that we combine item association relationships with sequential relationships in cascading style for sequential user behavior. 

\begin{figure*}[h!]
%第一行 loss
\centering
\begin{minipage}[t]{0.3\textwidth}
\centering
\includegraphics[width=\textwidth]{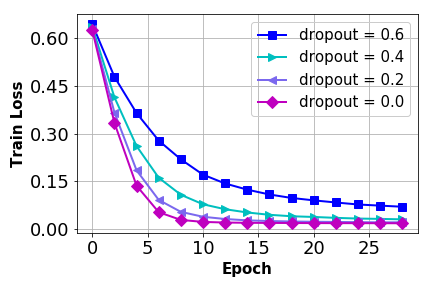}
\vspace{-22pt}
\caption*{(a) AIV - Train BPR Loss}
\end{minipage}
\begin{minipage}[t]{0.3\textwidth}
\centering
\includegraphics[width=\textwidth]{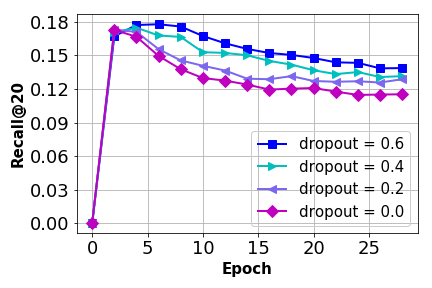}
\vspace{-22pt}
\caption*{(b) AIV - Recall@20}
\end{minipage}
\begin{minipage}[t]{0.3\textwidth}
\centering
\includegraphics[width=\textwidth]{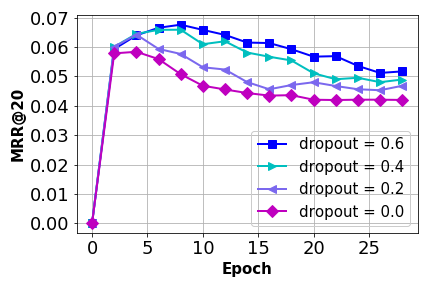}
\vspace{-22pt}
\caption*{(c) AIV - MRR@20}
\end{minipage}
%第二行 recall@20
\begin{minipage}[t]{0.3\textwidth}
\centering
\includegraphics[width=\textwidth]{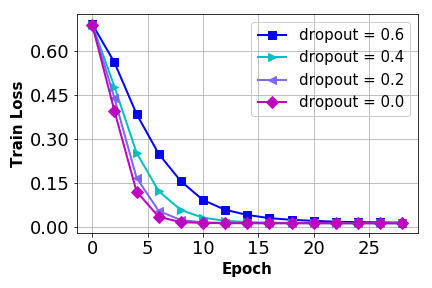}
\vspace{-22pt}
\caption*{(d) ACPA - Train BPR Loss}
\end{minipage}
\begin{minipage}[t]{0.3\textwidth}
\centering
\includegraphics[width=\textwidth]{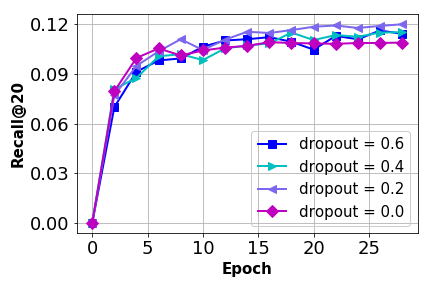}
\vspace{-22pt}
\caption*{(e) ACPA - Recall@20}
\end{minipage}
\begin{minipage}[t]{0.3\textwidth}
\centering
\includegraphics[width=\textwidth]{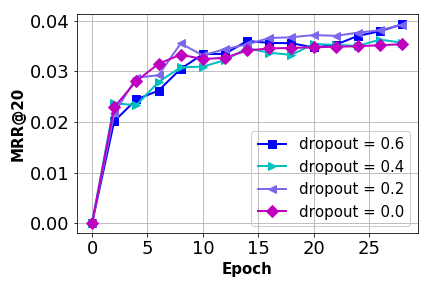}
\vspace{-22pt}
\caption*{(f) ACPA - MRR@20}
\end{minipage}

%第三行 mrr@20
\begin{minipage}[t]{0.3\textwidth}
\centering
\includegraphics[width=\textwidth]{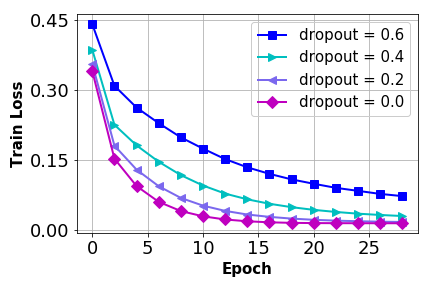}
\vspace{-22pt}
\caption*{(g) TaoBao - Train BPR Loss}
\end{minipage}
\begin{minipage}[t]{0.3\textwidth}
\centering
\includegraphics[width=\textwidth]{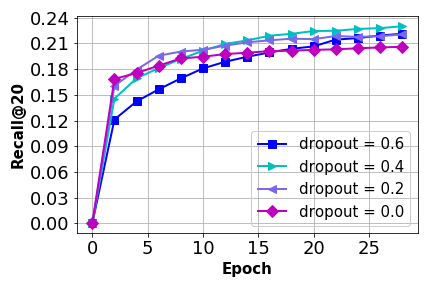}
\vspace{-22pt}
\caption*{(h) TaoBao - Recall@20}
\end{minipage}
\begin{minipage}[t]{0.3\textwidth}
\centering
\includegraphics[width=\textwidth]{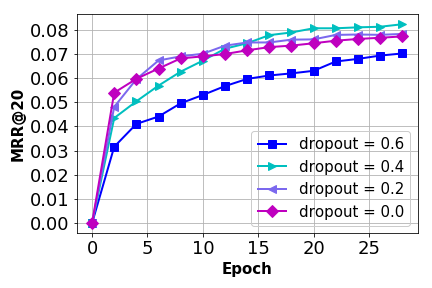}
\vspace{-22pt}
\caption*{(i) TaoBao - MRR@20}
\end{minipage}
\caption{Train loss and performance of Recall@20 and MRR@20 with different Dropout ratio on three datasets of CAASR. Dropout=0.0 means the Dropout keep probability equals 1.0. (a)(b)(c): train BPR loss on three datasets; (d)(e)(f): Recall@20 on three datasets; (g)(h)(i): MRR@20 on three datasets.}
\label{figure:performance_dropout}
\end{figure*}
\subsubsection{\textbf{Dropout Regularization.}}
In our method, we apply Dropout to play as the regularization technique for RNNs based methods. In order to study the influence of Dropout, we explore the model performance with different Dropout ratios along the training process on three datasets and the results are shown in Figure~\ref{figure:performance_dropout}. 

From the figure, we can see that different Dropout may have different influence on different datasets. In Figure~\ref{figure:performance_dropout} (a)(d)(g) for three datasets, it is obvious that no dropout (Dropout=0.0) will encourage the model to obtain lower loss, yet fit the train data better. While the model performance varies a lot with Dropout=0.0 for three datasets. For instance, in Figure~\ref{figure:performance_dropout} (e)(f) for ACPA and (h)(i) for TaoBao, CAASR can still reach a not bad performance when Dropout rate equals 0.0, which means no Dropout technique is utilized. This mainly due to that the data distribution is complex and no Dropout may encourage the model to fit data better. In contrary, no Dropout technique is not suitable for AIV dataset in Figure~\ref{figure:performance_dropout} (b)(c), because it is clearly that CAASR with no Dropout over-fits the data a lot. And Dropout rate equals 0.6 for AIV presents better performance, which indicates CAASR needs some regularization to avoid over-fitting. In summary, the Dropout parameter is better chosen according to specific dataset in order to robustly match the data distribution. 
\subsection{Qualitative Results}
\begin{figure*}[h!]
\centering
\begin{minipage}[t]{0.48\textwidth}
\centering
\includegraphics[width=\textwidth]{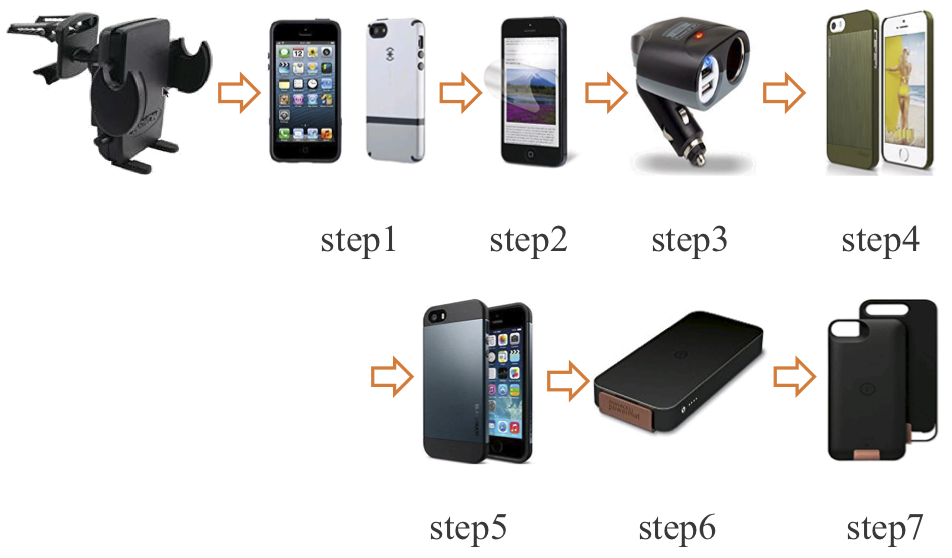}
\caption*{(a) user1 sequence on ACPA}
\end{minipage}
\begin{minipage}[t]{0.48\textwidth}
\centering
\includegraphics[width=\textwidth]{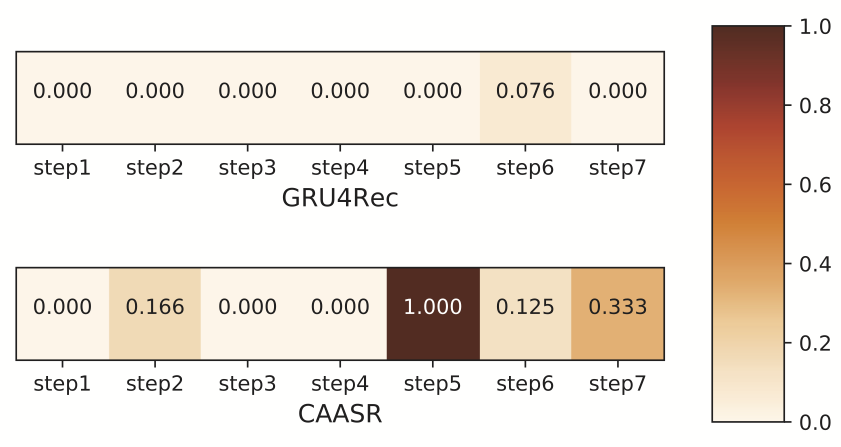}
\caption*{(b) predictions for user1 at each step on ACPA}
\end{minipage}

\vspace{0.4cm}

\begin{minipage}[t]{0.48\textwidth}
\centering
\includegraphics[width=\textwidth]{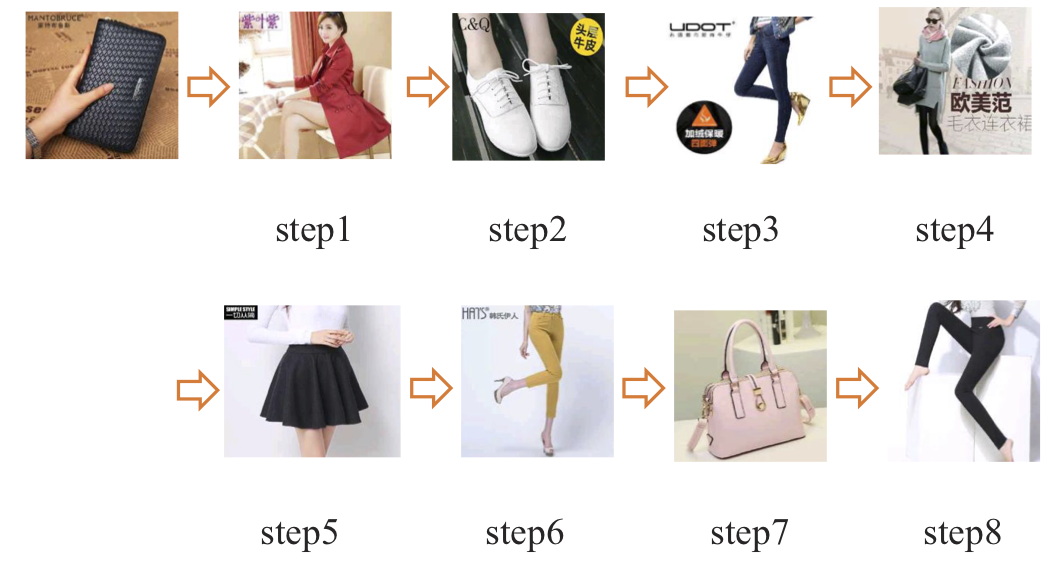}
\caption*{(c) user2 sequence on TaoBao}
\end{minipage}
\begin{minipage}[t]{0.48\textwidth}
\centering
\includegraphics[width=\textwidth]{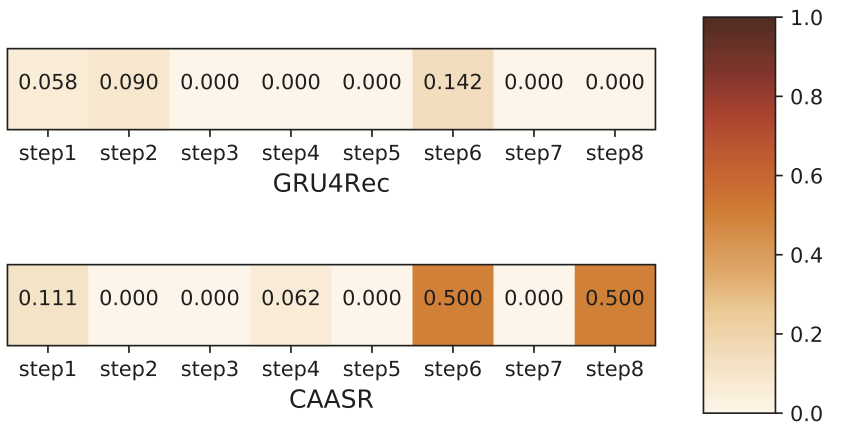}
\caption*{(d) predictions for user2 at each step on TaoBao}
\end{minipage}
\caption{Examples of sequence recommendation between GRU4Rec and our CAASR on ACPA and TaoBao datasets. Predictions at each step are listed and the MRR@20 values are tagged as well. Deeper orange means better MRR@20. And 0 denotes that model does not correctly predict the right one in top20.}
\label{figure:examples_recommendation}
% \vspace{-2pt}
\end{figure*}
\subsubsection{\textbf{Sequential Recommendation.}} Remind that our CAASR method integrally considers the item association relationships and sequential relationships for recommendation in sequential user behavior. While the typical RNNs based sequential recommendation method GRU4Rec mainly focus on modeling the sequential relationships. Therefore, it is curious to see whether CAASR benefits from item association in real recommendation sequence and we conduct an experiment to verify about this. In particular, Figure~\ref{figure:examples_recommendation}~(a)(c) show two examples in test set on ACPA and TaoBao datasets respectively. The corresponding predictions of MRR@20 at each step are stated in Figure~\ref{figure:examples_recommendation}~(b) and Figure~\ref{figure:examples_recommendation}~(d) respectively.  

ACPA is a dataset of user consuming history on cell phones and accessories. Sequence items in user1's history are shown in Figure~\ref{figure:examples_recommendation}~(a). Compared to cell phone in step 1 and its accessories at the following stpng, phone screen savers at step 2, vehicle charger at step 3, phone cases at step 4,5,7 and mobile source at step 6, these items do not have obvious sequential relationships, they present more kind of association relationships. We notice that in Figure~\ref{figure:examples_recommendation}~(b), our model CAASR exactly predicts the phone sticker at step 2 after the cell phone consumption at step 1, while GRU4Rec fails. This means our model is capable of capturing item association relationships that GRU4Rec misses. At step 4 and step 5, items both are phone shells. Regarding item association relationships among items, our model exactly predicts the phone shell at step 5 and puts it at the first place. This is very practical in real scenarios. 

Items in TaoBao dataset take the records of user purchase history on fashion and clothing. From the consuming sequence of user2 in Figure~\ref{figure:examples_recommendation}~(c), we notice that the user was inclined to buy some warm clothes like at step 1, step 3 and step 4 mainly due to the whether. From step 4 to step 5, the user began to buy some cool clothing. In fact, change from warm clothes to cool ones indeed exists sequential relationships because of the change of seasons. While in one particular season or whether, like items at stpng 1,2,3,4 and items at stpng 5,6,7,8, there is no particular sequential relationships. We note that the pale brown trousers at step 6 and black trousers at step 8 are more like association relationships rather than sequential relationships. Therefore in Figure~\ref{figure:examples_recommendation}~(d), we can see that our CAASR, involved with this kind of association relationships, achieves a better performance than GRU4Rec at these stpng.

\begin{figure*}[!htb]
\centering
\subfloat[]{\includegraphics[width = 0.24\linewidth]{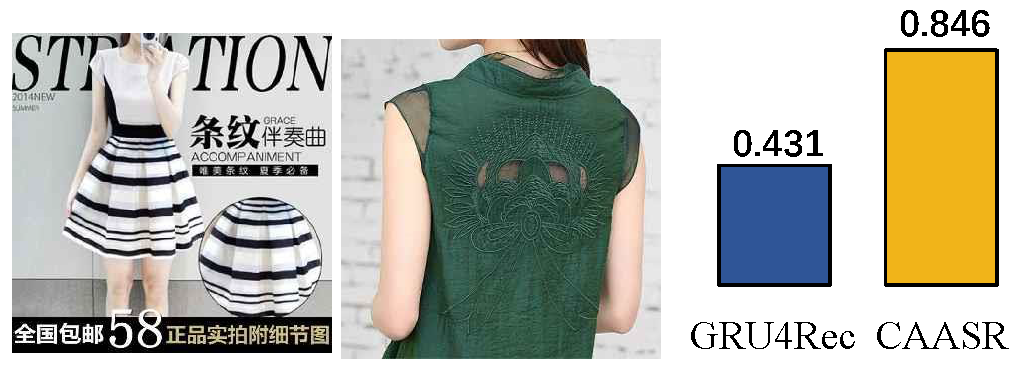}}\
\subfloat[]{\includegraphics[width = 0.24\linewidth]{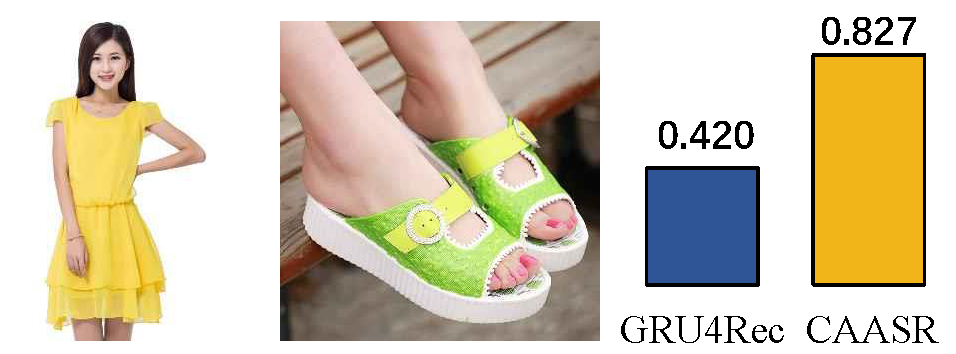}}\
\subfloat[]{\includegraphics[width = 0.24\linewidth]{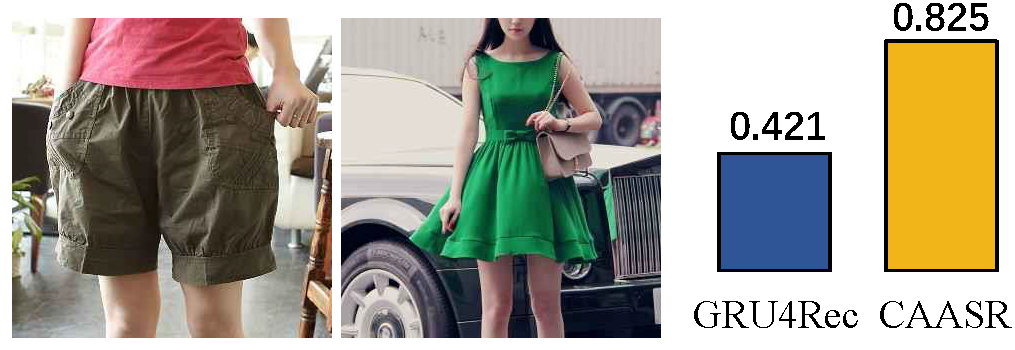}}\
\subfloat[]{\includegraphics[width = 0.24\linewidth]{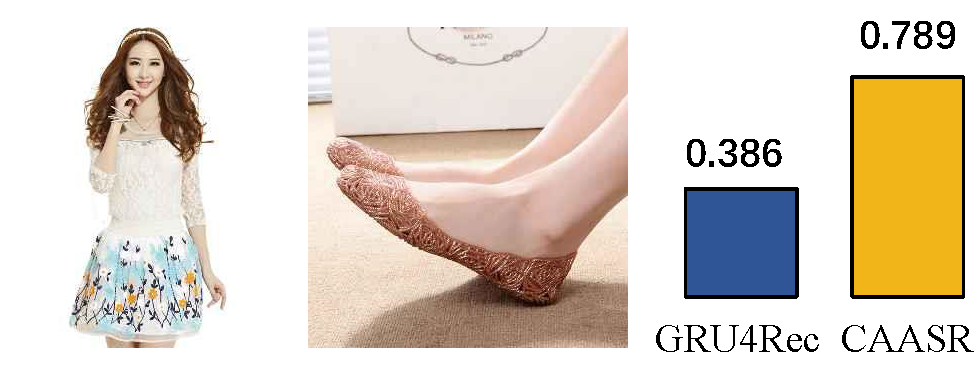}}\\
\subfloat[]{\includegraphics[width = 0.24\linewidth]{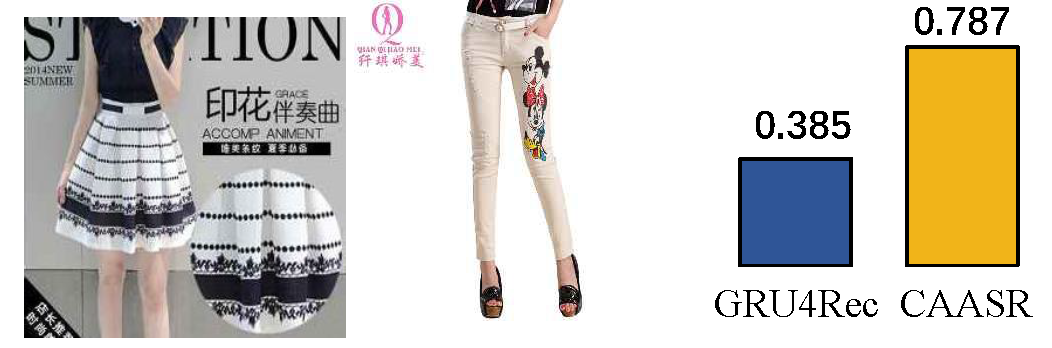}}\
\subfloat[]{\includegraphics[width = 0.24\linewidth]{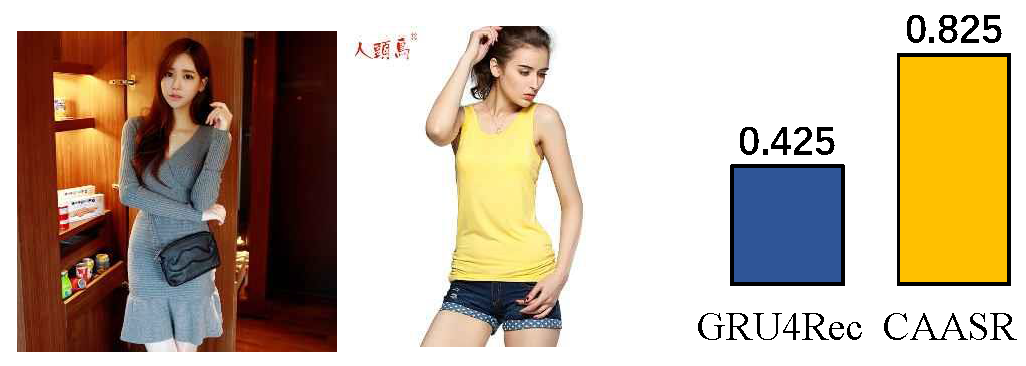}}\
\subfloat[]{\includegraphics[width = 0.24\linewidth]{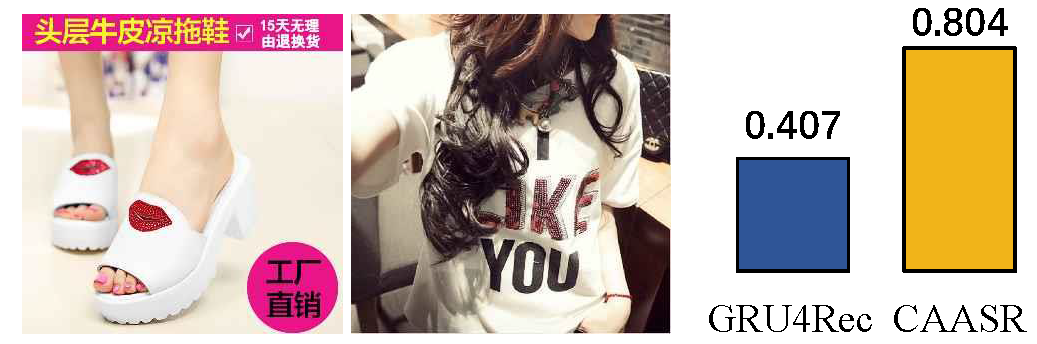}}\
\subfloat[]{\includegraphics[width = 0.24\linewidth]{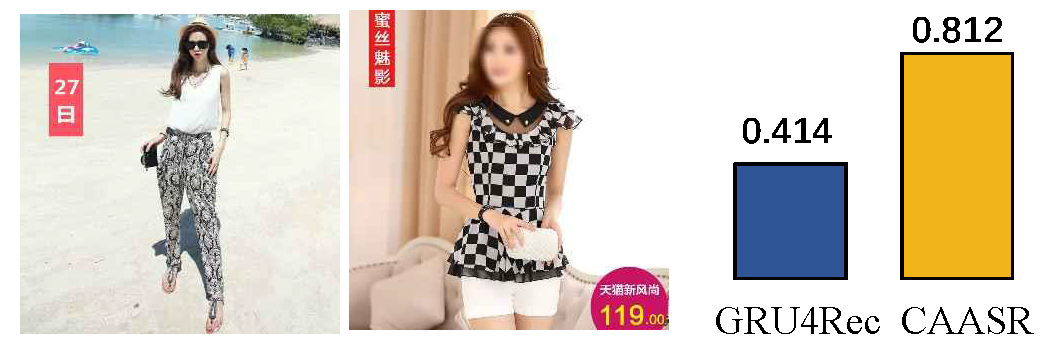}}\\
\subfloat[]{\includegraphics[width = 0.24\linewidth]{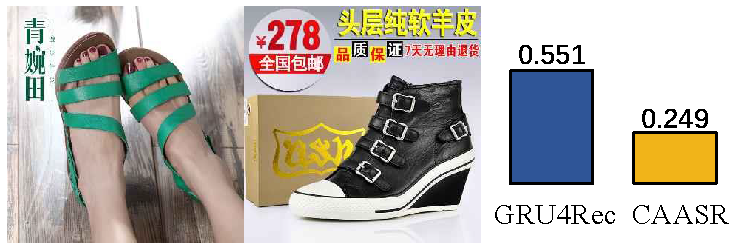}}\
\subfloat[]{\includegraphics[width = 0.24\linewidth]{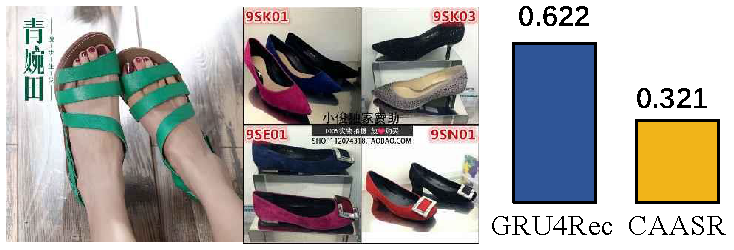}}\
\subfloat[]{\includegraphics[width = 0.24\linewidth]{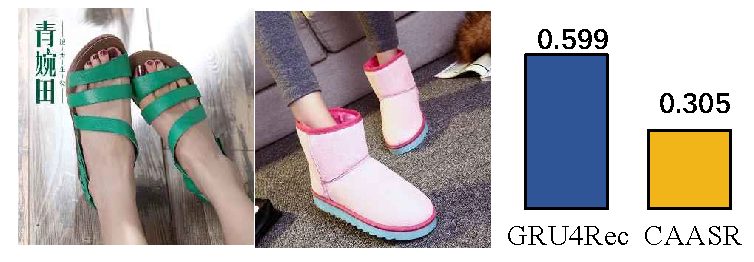}}\
\subfloat[]{\includegraphics[width = 0.24\linewidth]{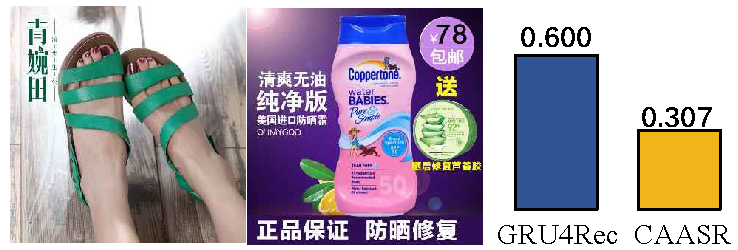}}\\
\subfloat[]{\includegraphics[width = 0.24\linewidth]{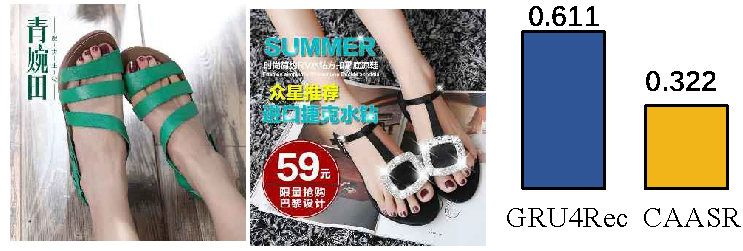}}\
\subfloat[]{\includegraphics[width = 0.24\linewidth]{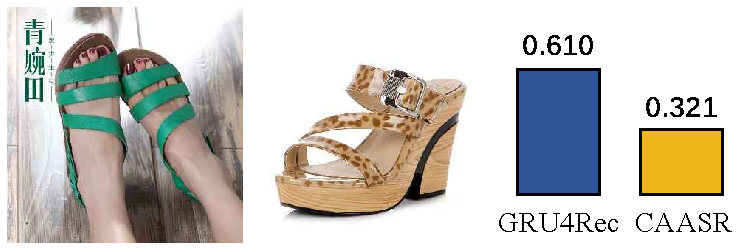}}\
\subfloat[]{\includegraphics[width = 0.24\linewidth]{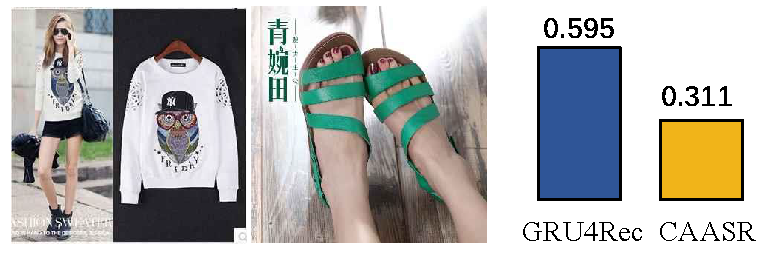}}\
\subfloat[]{\includegraphics[width = 0.24\linewidth]{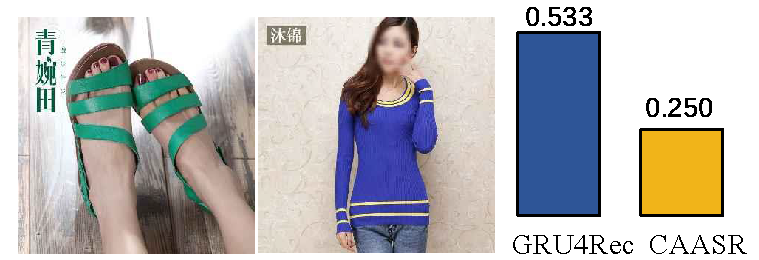}}\\
\caption{Comparison of cosine similarity based on two types of item embeddings on TaoBao dataset. The blue bar and yellow bar represent cosine similarity based on GRU4Rec item embedding and CAASR item embedding respectively. Item pairs with top 8 largest differences in the two types of cosine similarity are shown.  (a-h): CAASR is with higher similarity. (i-p): GRU4Rec is with higher similarity.}
\label{figure:similarity_analysis}
\end{figure*}
\subsubsection{\textbf{Item Embedding Analysis.}}
Both GRU4Rec and our model attempt to learn item embeddings from user-item interaction data. Similar item embeddings are expected to share similar properties in terms of user-item interaction. However, GRU4Rec mainly leverages sequential relationships to learn item representation, whereas our model integrates both the sequential relationships and association relationships. Therefore, our model is expected to better capture the item association relationships compared to GRU4Rec. 

To verify the above hypothesis, we calculate the pair-wise item cosine similarity for all item pairs based on learned item embeddings. Let $S^1_{ij} (i,j=1,\cdots,N)$ and $S^2_{ij} (i,j=1,\cdots,N)$ denote the cosine similarity of items $i$ and $j$ based on item embeddings for GRU4Rec and CAASR respectively. We take 8 item pairs with the largest difference $S^+=S^2-S^1$. The corresponding item images are shown in Figure~\ref{figure:similarity_analysis}~(a-h). We can see that the item pairs are mainly complementary products like skirt with sandals in Figure~\ref{figure:similarity_analysis}~(b)(d). This suggests that compared to GRU4Rec embeddings, our CAASR model can better capture the association relationships among complementary products. 

We also take 8 item pairs with the largest difference $S^-=S^1 - S^2$. The corresponding item images are shown in Figure~\ref{figure:similarity_analysis}~(i-p). In this figure, we see that item pairs mainly focus on items in same category like shoes in Figure~\ref{figure:similarity_analysis}~(i)(j) or irrelevant item pairs like shoes with washing liquid in Figure~\ref{figure:similarity_analysis}~(l) and sandals with snow boots in Figure~\ref{figure:similarity_analysis}~(k). This suggests that compared to CAASR embedding, GRU4Rec captures relatively pointless association relationships among products.

This matches well with our expectation as we introduce the item association graph to model the association relationships in sequential user behavior. Our model is able to recommend the associated and complementary items. This characteristic benefits the recommendation accuracy in real scenarios because a user tends to buy an associated or complementary item at next step rather than some irrelevant products.

\section{Conclusion and Future Work}
In this paper, we point that both the association relationships and sequential relationships are existed in sequential user behavior data. These two inherent characteristics make it essential to capture both of them for sequential recommendation. Although RNNs based sequential recommendation methods are capable of model sequential relationships, they fail to put emphasize on association relationships. In this case, we propose a cascading CAASR model that unifiedly incorporates item association relationships and sequential relationships for sequential recommendation in cascading style. In addition, two parallel styles of combining this item association relationships and sequential relationships are explored and analyzed in this paper to demonstrate the advantages of cascading style. To the best of our knowledge, CAASR is the first model that concentrates on item association relationships in RNNs based sequential recommendation. We conduct extensive experiments on three real-world datasets: AIV, ACPA and TaoBao. Results demonstrate the effectiveness of our model both quantitatively and qualitatively.

In future, we will explore one more general and efficient framework to combine the association relationships and sequential relationships for modeling sequential user behavior and make better recommendation performance in this scenario.

\begin{acks}

\end{acks}

\newpage
% Bibliography
\bibliographystyle{ACM-Reference-Format}
\bibliography{main}

\end{document}